\documentclass[aps,prc,twocolumn,showpacs,preprintnumbers,superscriptaddress,floatfix,nofootinbib,amsmath,amssymb,balancelastpage,table]{revtex4-1}
\usepackage{bm}
\usepackage{amsmath}
\usepackage{multirow}
\usepackage{ifpdf}
\usepackage{url}

\ifpdf
  \usepackage{graphicx}     %   usepackage without driver option
  \usepackage{hyperref}
\else     % For (p)LaTeX + dvipdfmx
  \usepackage[dvipdfmx]{graphicx}     %   usepackage with driver option
  \usepackage[dvipdfmx]{hyperref}
\fi
\usepackage{color}

\definecolor{Tomato1}{rgb}{1,0.39, 0.28}
\definecolor{Tomato2}{rgb}{0.932,0.36, 0.26}
\definecolor{Tomato3}{rgb}{0.804,0.31, 0.224}
\definecolor{Tomato4}{rgb}{0.545,0.21, 0.15}

\definecolor{RoyalBlue1}{rgb}{ 0.284, 0.464, 1}
\definecolor{RoyalBlue2}{rgb}{  0.264, 0.43,  0.932 }
\definecolor{RoyalBlue3}{rgb}{  0.228, 0.372, 0.804 }
\definecolor{RoyalBlue4}{rgb}{  0.152, 0.25,  0.545}

\hypersetup{
colorlinks=true,
linkcolor=RoyalBlue2,
citecolor=Tomato2,
urlcolor=Tomato2
}

\newcommand\pythia{\textsc{Pythia}}
\newcommand\fastjet{\texttt{FastJet}}

\renewcommand{\epsilon}{\varepsilon}

\newcommand{\BV[1]}{\mbox{\boldmath ${#1}$}}

\begin{document}
\title{Full jet in quark-gluon plasma with hydrodynamic medium response}

\author{Yasuki~Tachibana}\email{yasuki.tachibana@mail.ccnu.edu.cn}
\affiliation{Institute of Particle Physics and Key Laboratory of Quark and Lepton Physics (MOE), Central China Normal University, Wuhan, Hubei, 430079, China}
\author{Ning-Bo~Chang}
\email{changnb@mail.ccnu.edu.cn}
\affiliation{Institute of Theoretical Physics, Xinyang Normal University, Xinyang, Henan 464000, China}
\affiliation{Institute of Particle Physics and Key Laboratory of Quark and Lepton Physics (MOE), Central China Normal University, Wuhan, Hubei, 430079, China}
\author{Guang-You~Qin}
\email{guangyou.qin@mail.ccnu.edu.cn}
\affiliation{Institute of Particle Physics and Key Laboratory of Quark and Lepton Physics (MOE), Central China Normal University, Wuhan, Hubei, 430079, China}

\date{\today}

\begin{abstract}

We study the nuclear modifications of full jets and their structures in relativistic heavy-ion collisions including the effect of hydrodynamic medium response to jet quenching.
To study the evolutions of the full jet shower and the traversed medium with energy and momentum exchanges between them,
we formulate a coupled jet-fluid model consisting of a set of jet transport equations and relativistic hydrodynamics equations with source terms.
In our model, the full jet shower interacts with the medium and gets modified via collisional and radiative processes during the propagation.
Meanwhile, the energy and momentum are deposited from the jet shower to the medium and then evolve with the medium hydrodynamically.
The full jet defined by a cone size in the final state includes the jet shower and the particles produced from jet-induced flow.
We apply our model to calculate the full jet energy loss and the nuclear modifications of jet rate and shape in Pb+Pb collisions at $2.76{\rm A~TeV}$.
It is found that the inclusion of jet-induced flow contribution leads to stronger jet-cone size dependence for jet energy loss and jet suppression.
Jet-induced flow also has a significant contribution to jet shape function and dominates at large angles away from the jet axis.

\end{abstract}
%\pacs{12.38.Bx, 12.38.Mh, 13.87.Ce, 24.10.Nz, 24.85.+p, 25.75.-q, 25.75.Bh, 25.75.Ld}
\maketitle

\section{Introduction}
In relativistic heavy-ion collisions at the Relativistic Heavy Ion Collider (RHIC) and the Large Hadron Collider (LHC), the deconfined state of quarks and gluons, namely quark-gluon plasma (QGP), has been created.
One of the remarkable properties of the produced QGP is the strong collective motion which has been well described by relativistic hydrodynamics with the extremely small viscosity to entropy ratio \cite{Heinz:2001xi, sQGP1, sQGP2, sQGP3, Hirano:2005wx, Schenke:2010rr, Qiu:2011iv, Qiu:2011hf, Gale:2012rq}.
The hydrodynamic behavior of the QGP has been confirmed by the large anisotropic flows as measured by experiments, which implies strong interactions among the QGP constituents.

In addition to the collective phenomena, one can study the novel features of the QGP as strongly interacting matter via jet quenching \cite{Bjorken:1982tu, Appel:1985dq, Blaizot:1986ma, Rammerstorfer:1990js, Gyulassy:1990ye, Thoma:1990fm, Gyulassy:1993hr, Qin:2015srf, Blaizot:2015lma}.
In high-energy nucleus-nucleus collisions, the energetic jet showering partons may be produced in hard processes at very early time.
During their propagation through the QGP, the jet partons interact with the medium constituents via collisional and radiative processes, which change the momenta of the shower partons in the hard jet. The phenomena related to the modification of the jets and their inner structure by the QGP medium effect are commonly referred to as jet quenching in a broad sense.

In recent experiments of Pb+Pb collisions at the LHC, detailed measurements of fully reconstructed jets with very high transverse momentum have become available, thanks to the high center of mass energy.
One of the most general consequence of jet quenching seen in the experiments is the suppression of jet rates due to the energy loss of the full jets with a given cone of the size $R = \sqrt{\Delta \eta_p^2 + \Delta\phi_p^2}$ \cite{Aad:2012vca, Abelev:2013kqa, Aad:2014bxa, Adam:2015ewa,Khachatryan:2016jfl}.
Furthermore, the measurements of substructures inside the full jets have provided us more detailed information on jet quenching \cite{Chatrchyan:2013kwa, Chatrchyan:2014ava, Aad:2014wha, Khachatryan:2015lha, CMS:2015kca, ATLAS:2015aa, CMS:2016jys}.
Motivated by the detailed measurements, a lot of theoretical effort has been devoted to the study of jet-medium interaction and the medium effect on the full jets with showering structures in relativistic heavy-ion collisions
\cite{Vitev:2009rd,CasalderreySolana:2010eh,Qin:2010mn,MehtarTani:2011tz,Lokhtin:2011qq,Young:2011qx,He:2011pd, Renk:2012cx, Dai:2012am,Qin:2012gp,CasalderreySolana:2012ef,Apolinario:2012cg,Zapp:2012ak,Majumder:2013re, Blaizot:2013hx,Wang:2013cia, Ma:2013pha,Senzel:2013dta,Ramos:2014mba, Casalderrey-Solana:2014bpa, Blaizot:2014ula,Fister:2014zxa,Perez-Ramos:2014mna,He:2015pra,Chien:2015hda, Milhano:2015mng,Chang:2016gjp, Mueller:2016gko, Chen:2016vem,Casalderrey-Solana:2016jvj,Mehtar-Tani:2016aco,Wang:2016fds,Gao:2016ldo,Chen:2016cof}

The interactions between the energetic partons in the jet shower and the QGP medium constituents exchange energies and momenta between them.
For the QGP fluid, the jet shower is a bunch of fast-moving energy-momentum deposition sources, which are supposed to excite the medium fluid, and induce flows propagating with the jet as a collective medium response \cite{Stoecker:2004qu, CasalderreySolana:2004qm,Ruppert:2005uz, Satarov:2005mv, Renk:2005si,Ma:2006fm, Chaudhuri:2006qk, Zhang:2007qx, Chaudhuri:2007vc, Gubser:2007ga, Chesler:2007sv,Neufeld:2008fi, Neufeld:2008hs, Noronha:2008un, Neufeld:2008dx, Neufeld:2009ep,Qin:2009uh, Li:2009ax, Neufeld:2010tz, Betz:2010qh, Ma:2010dv, Neufeld:2011yh,Bouras:2012mh,
Bouras:2014rea, Tachibana:2014lja, Andrade:2014swa, Floerchinger:2014yqa, Schulc:2014jma, He:2015pra, Tachibana:2015qxa}.
The jet-induced flow carries the energy and momentum deposited by the jet and enhances the hadron emission from the medium around the direction of the jet axis.
Some of these enhanced hadrons are detected as part of the jet together with the fragments from the jet shower and affect the final state full jet energy and structures \cite{He:2015pra,Casalderrey-Solana:2016jvj,Wang:2016fds}.
Since the structure of the jet-induced flow is characterized by the bulk properties of the QGP, e.g., sound velocity, and viscosities, the detailed investigations of the contribution of the collective medium response to the jet structure may provide us not only the precise interpretation of the experimental data, but also unique opportunities to study the bulk properties of the QGP through the jet events in relativistic heavy-ion collisions.

In this work, we study the nuclear modification of full jet structures in the QGP medium including the contributions from the hydrodynamic medium response to jet-deposited energy and momentum.
We employ a coupled full jet shower and QGP fluid model which is composed of a set of transport equations for the jet shower evolution and the hydrodynamic equations with source terms for the QGP medium evolution.
The transport equations describe the evolution of the three-dimensional momentum distributions of partons in the jet shower \cite{Chang:2016gjp}, including the collisional energy loss, the transverse momentum broadening, and medium-induced partonic splittings for all partons within the showering jet.
The space-time evolution of the QGP medium is described by (3+1)-dimensional relativistic ideal hydrodynamic equations with source terms \cite{Tachibana:2014lja,Tachibana:2015qxa}.
The source terms account for the transfer of the deposited energy and momentum from the jet shower to the QGP fluid, and are constructed with the evolving distributions of the partons in the jet shower
obtained as solutions of the jet transport equations.
Based on our coupled jet-fluid model, we perform simulations of jet events in Pb+Pb collisions at 2.76{\rm A~TeV}, investigate the flow induced in the medium as a hydrodynamic response to jet quenching, and study how the jet-induced flow affects the full jet structures in the final state.
Our study shows that the contribution of the particles originating from the jet-induced flows increases the jet-cone size dependence of the full jet energy loss.
It is also found that jet-induced flow has a significant contribution to the final state full jet shape, and dominates jet shape function at very large angles away from the jet direction.

The paper is organized as follows. In Sec.~\ref{sec:model}, we present the formulation of our coupled full jet shower and QGP fluid model used in this work.
In Sec.~\ref{sec:res}, we present and discuss the results from the simulations of jet events in Pb+Pb collisions at 2.76{\rm A~TeV}.
We will focus on the effect of the hydrodynamic medium response to the jet quenching on the final state full jet observables. Section~\ref{sec:sum} is devoted to the summary and concluding remarks of this work.

%Hereafter we use units such that $\hbar=c=k_{\rm B} =1$. The collision axis is chosen as the $z$-axis and the incident Pb nuclei collide at $(t,\,z)=(0,\,0)$.
%We also use the relativistic $\tau$-$\eta_{\rm s}$ coordinates $\left(\tau, x, y, \eta_{\rm s}\right)$, where $\tau=\left(t^2-z^2\right)^{1/2}$ is the proper time, and $\eta_{\rm s}=\left(1/2\right)\ln\left[\left(t+z\right)/\left(t-z\right)\right]$ is the space-time rapidity.
%We denote the pseudorapidity as $\eta_{p}=\left(1/2\right)\ln\left[\left(E+p_z\right)/\left(E-p_z\right)\right]$, the azimuthal angle in the configuration space as $\phi=\arctan(y/x)$, and that in the momentum space as $\phi_{p}=\arctan(p_y/p_x)$.

\section{The Coupled Jet-Fluid Model}\label{sec:model}

\subsection{Jet Shower Evolution in Medium}\label{shower}

Jets are collimated clusters of the particles originating from high-$p_T$ partons produced in early stage partonic hard scatterings.
The produced high-$p_T$ parton successively radiates partons and develops a shower of partons due to its high virtuality.
In relativistic heavy-ion collisions, the interactions between the propagating jet and the constituents of the QGP medium, including collsional and radiative processes, change
the momenta of the jet shower partons, and thus modify the energy
as well as the structure of the full jet defined by a cone size.

In this work, to describe the time-evolution of the jet shower structure driven by the interaction with the medium, we employ a set of coupled transport equations for the energy and transverse momentum distributions of the partons contained in the jet shower,
\begin{eqnarray}
  \label{eq:jetdist}
f_i(\omega_i, k_{i\perp}^2,t)=\frac{dN_i(\omega_i, k_{i\perp}^2,t)}{d\omega_i dk_{i\perp}^2},
\end{eqnarray}
where the index $i$ denotes the parton species (quark or gluon), $\omega_i$ is its energy, and $k_{i\perp}$ is its transverse momentum with respect to the jet axis.
The transport equations have the following generic form \cite{Chang:2016gjp}:
\begin{eqnarray}
  \label{eq:jetshower}
  \!\!&&\!\!\frac{d}{dt}{f}_j(\omega_j, k_{j\perp}^2, t) = \left(\hat{e}_j \frac{\partial}{\partial \omega_j}
  + \frac{1}{4} \hat{q}_j {\nabla_{k_\perp}^2}\right){f}_j(\omega_j, k_{j\perp}^2, t)  \ \ \
  \nonumber \\
&&\!\!+\sum_i\int d\omega_idk_{i\perp}^2 \frac{d\tilde{\Gamma}_{i\rightarrow j}(\omega_j, k_{j\perp}^2|\omega_i, k_{i\perp}^2)}{d\omega_j dk^2_{j\perp}dt} f_i(\omega_i, k_{i\perp}^2, t)\nonumber\\
  \!\!&&\!\!-\sum_i \int d\omega_idk_{i\perp}^2 \frac{d\tilde{\Gamma}_{j\rightarrow i}(\omega_i, k_{i\perp}^2|\omega_j, k_{j\perp}^2)}{d\omega_i dk^2_{i\perp}dt} {f}_j(\omega_j, k_{j\perp}^2, t).
\end{eqnarray}
The first two terms on the right hand side in Eq.~(\ref{eq:jetshower}) describe the energy-momentum changes of the jet partons via
scatterings with the medium constituents.
The first term accounts for the energy-momentum changes in the longitudinal direction with respect to the jet axis; such contribution is called collisional energy loss and its strength is determined by the longitudinal momentum loss rate $\hat{e}=d\langle E\rangle/dt$.
The momentum changes in the transverse direction, whose effect is called transverse momentum broadening, is taken into account through the second term, with the exchange rate of transverse momentum squared $\hat{q}=d\langle \Delta p_\perp^2\rangle/dt$ \cite{Baier:1996kr, Majumder:2007hx, Qin:2012fua}.
The last two terms are the contributions from the medium-induced partonic splitting processes, including the gain term coming from the radiation of parton $j$ with $\omega_j$ and $k_{j\perp}$ from the parton $i$ with $\omega_i$ and $k_{i\perp}$ and the loss term for the radiation of parton $i$ from parton $j$.
It should be noted that the transport equations~(\ref{eq:jetshower}) are coupled to each other through the terms for the medium-induced parton splittings, i.e., the gain term for the process $i\rightarrow j$ in the transport equation for parton $j$ also appears as a loss term in the transport equation for parton $i$.
The medium-induced parton splittings change the number of partons and the internal momentum distribution of the jet, but do not change the total energy and momentum inside the full jet (for sufficiently large cone size) because they are conserved in the splitting processes.

For the rates of the medium-induced partonic splittings, we employ the results from higher-twist jet energy loss formalism \cite{Wang:2001ifa, Majumder:2009ge}:
\begin{eqnarray}
\!\!\!\!\!\!\frac{d\Gamma_{i\rightarrow j}(\omega_j, k_{j\perp}^2|E_i)} {d\omega_j dk_{j\perp}^2 dt} = \frac{2 \alpha_s}{\pi}\hat{q}_g \frac{ x P_{i\rightarrow j}(x)  }{\omega_j k_{j\perp}^4} \!\sin^2 \!\left(\frac{t - t_{i}}{2\tau_f}\right).
  \label{eq:radiation}
\end{eqnarray}
Here, $P_{i\rightarrow j}(x=\omega_j/E_i)$ is the vacuum splitting function for the process $i\rightarrow j$ with $\omega_j$ the energy of the radiated parton and $E_i$ the energy of the parent parton, $\tau_f = 2E_ix(1-x)/k_{j\perp}^2$ is the formation time of the radiated parton with $k_{j\perp}$ the transverse momentum with respect to the propagation direction of the parent parton, and $t_{i}$ is the production time of the parent parton.
To obtain the splitting rates of the form $\frac{d\tilde{\Gamma}_{i\rightarrow j}(\omega_j, k_{j\perp}^2|\omega_i, k_{i\perp}^2)}{d\omega_j d^2k_{j\perp}dt}$
in Eq.~(\ref{eq:jetshower}), the multiplication with the Jacobian $\mathcal{J} = \left|\frac{\partial k_{ij\perp}^2}{\partial k_{j\perp}^2}\right|$ is necessary (see Ref. \cite{Chang:2016gjp} for more details).
In the calculation, we put a constraint that the medium-induced radiations are allowed only for the partons whose formation times are achieved ($t>\tau_f$).

In our model, the strength of the each contribution of the medium modification in Eq.~(\ref{eq:jetshower}) is controlled by two transport coefficients: $\hat{e}$ for collisional energy loss,
and $\hat{q}$ for transverse momentum broadening and medium-induced partonic splittings.
Assuming that the medium is in almost local thermal equilibrium, the fluctuation-dissipation theorem can relate these two transport coefficients to each other: $\hat{q} = 4T \hat{e}$ \cite{Moore:2004tg, Qin:2009gw}.
Also since $\hat{q}$ for quarks and gluons are related by a color factor, $\hat{q}_g/\hat{q}_q = C_{\rm A}/C_{\rm F}$, only one transport coefficient (we choose $\hat{q}_{q}$ for quark) determines the sizes of all the medium effects in Eq.~(\ref{eq:jetshower}).
In this work, we use $\hat{q}_q$ of the form,
\begin{equation}
\label{eq:qhat}
\hat{q}_q (x) = \hat{q}_{q,0} \left[\frac{T(x)}{T_{0}}\right]^3  \frac{p\cdot u(x)}{p_0}.
\end{equation}
where $T_0=T\left(\tau=\tau_0,x=0,y=0,\eta_{\rm s}=0\right)$ is the initial temperature at the center of the QGP medium produced in central Pb$+$Pb collisions at 2.76{\rm A~TeV} at the LHC, $p^\mu$ is the four-momentum of the propagating parton, $u_\mu$ is the flow four-velocity of the medium, and $\hat{q}_{q,0}$ is the exchange rate of transverse momentum squared for quark in a static QGP medium at $T_0$. The factor ${p\cdot u}/{p_0}$ is introduced to account for the flow effect in a non-static medium \cite{Baier:2006pt}.

We solve the transport equations~(\ref{eq:jetshower}) numerically to describe the jet evolution in terms of the energy and transverse momentum distributions of the partons contained in the jet shower.
The initial condition for full jet shower is constructed from \pythia\:simulation \cite{Sjostrand:2007gs}. 
The jet shower evolves according to the transport equations 
during the propagation through the QGP medium with the local temperature 
above $T_c = 160~{\rm MeV}$.
We introduce a minimum energy $\omega_{\rm cut}$ for the partons in the jet shower.
When the energy of the partons in the jet shower becomes below $\omega_{\rm cut}$ during their evolution, they are considered to be absorbed in the medium.
We put the energy and momentum of the absorbed partons into the medium together with the collisional energy loss and transverse momentum broadening terms in Eq.~(\ref{eq:jetshower}).
The same cut-off energy is also used for the medium-induced partonic splittings, i.e, only partons with energy above $\omega_{\rm cut}$ can be radiated; this is to take into account some effect for the balance between radiation and absorption.
In this study, we set $\hat{q}_{q,0}=1.7~{\rm GeV^2/fm}$ and $\omega_{\rm cut}=1~{\rm GeV}$. These values provide a good description to the nuclear modification factor $R_{\rm AA}$ for inclusive jet spectrum in most central Pb+Pb collisions at $2.76{\rm A~TeV}$ measured by the ATLAS, ALICE, and CMS Collaborations \cite{Aad:2012vca, Abelev:2013kqa, Khachatryan:2016jfl}.
Our value of $\hat{q}_{q,0}$ is also consistent with the one obtained by JET Collaboration \cite{Burke:2013yra}.

%%%%%%%%%%%%%%%%%%%%%%%%%%%%%%%%%%%%%%%%%%
\subsection{Hydrodynamic Equations with Source Terms}

The conventional way to describe the space-time evolution of the QGP is to use the hydrodynamic equations $\partial_{\mu} T_{\rm QGP}^{\mu \nu}=0$,
where $T_{\rm QGP}^{\mu \nu}$ is the energy-momentum tensor of the QGP fluid.
This equation represents the energy-momentum conservation only in the fluid.
However, in the case that jets propagate through the QGP, the QGP and the jets exchange their energies and momenta through the scatterings between their constituents.
Therefore, the energy-momentum conservation is satisfied not for the QGP only, but for the combined system of the QGP and the jets:
\begin{eqnarray}
  \partial_{\mu} \left[T_{\rm QGP}^{\mu \nu}\!\!\left(x\right)+T_{\rm jet}^{\mu \nu}\!\!\left(x\right)\right]=0. \label{eq:emc}
\end{eqnarray}
Here
$T_{\rm jet}^{\mu \nu}$ is the energy-momentum tensor of the jet shower.
Assuming that the energy and momentum deposited by the jets are quickly thermalized, we model the QGP as an ideal fluid in local equilibrium whose energy-momentum tensor can be decomposed as:
\begin{eqnarray}
  T^{\mu\nu}_{\rm QGP}=\left(\epsilon+p\right)u^{\mu}u^{\nu}-p\eta^{\mu\nu},
\end{eqnarray}
where $\epsilon$ is the energy density, $p$ is the pressure, $u^{\mu}$ is the flow four-velocity, and $\eta^{\mu\nu}={\rm diag}\left(1,-1,-1,-1\right)$ is the Minkowski metric.
If we define the source terms by
\begin{eqnarray}
  \label{eq:st}
  J^{\nu}\!\!\left(x\right)=-\partial_{\mu}T_{\rm jet}^{\mu \nu}\!\!\left(x\right),
\end{eqnarray}
Equation (\ref{eq:emc}) becomes the hydrodynamic equations with source terms:
\begin{eqnarray}
  \partial_{\mu}T_{\rm QGP}^{\mu \nu}\!\!\left(x\right)&=&J^{\nu}\!\!\left(x\right). \label{eq:eom}
\end{eqnarray}
In the numerical hydrodynamic calculations for the medium evolution in the relativistic heavy-ion collisions, we use the relativistic $\tau$-$\eta_{\rm s}$ coordinates which is convenient to describe the longitudinal dynamics.
To do so, we employ the equations in which the partial derivative in the Eq.~(\ref{eq:eom}) are the replaced with the covariant derivative in the relativistic $\tau$-$\eta_{\rm s}$ coordinates:
\begin{eqnarray}
  D_{\bar\mu}T_{\rm QGP}^{\bar\mu \bar\nu}&=&J^{\bar\nu},
  \label{eq:covariant}
\end{eqnarray}
where $\bar\mu$ and $\bar\nu$ are the suffixes for the components in the $\tau$-$\eta_{\rm s}$ coordinate system: $x^{\bar\mu}=\left(\tau,x,y,\eta_{\rm s}\right)$.
Equation~(\ref{eq:covariant}) is numerically solved to describe the space-time evolution of the QGP medium.
Using this framework, the collective response to the jet quenching in the expanding QGP fluid can be properly described.

To close the system of hydrodynamic equations~(\ref{eq:eom}), an equation of state is necessary.
Here we employ the equation of state from the lattice {QCD} calculation \cite{Borsanyi:2013cga}.
As the fluid expands, the QGP fluid cools down and finally turns into a hadronic matter according to the equation of state.
We keep using Eq.~(\ref{eq:eom}) to describe the space-time evolution of the hadronic matter until the temperature drops to the freeze-out temperature $T_{\rm FO}$.

\subsection{The Source Terms}\label{source}

The source terms in Eq.~(\ref{eq:eom}) describe the transfer of the energy-momentum between the jet and the QGP fluid.
Here we construct the source terms from the evolving parton distributions in the jet shower obtained as the solutions of the transport equation~(\ref{eq:jetshower}).
We recall the kinetic definition of the energy-momentum tensor of the jet shower \cite{DeGroot:1980dk}:
\begin{eqnarray}
  T^{\mu \nu}_{\rm jet}\!\!\left(x\right)
  &=&
  \!\sum_{j}\int \frac{d^3k_j}{\omega_j} k_j^\nu k_j^\mu f_{j}\!\left(\BV[k]_j,\BV[x],t\right),
\end{eqnarray}
where $f_{j}\!\left(\BV[k]_j,\BV[x],t\right)$ is the phase-space distribution of the parton $j$ in the jet shower.
Referring to Eq.~(\ref{eq:st}), the source term can be written as:
\begin{eqnarray}
\!\!\!\!\!\!\!\!
J^{\nu}\!\!\left(x\right)
 =
 -\frac{dP_{\rm jet}^{{\nu}}}{dt d^3x}
 =
  -\!\sum_{j}\int \frac{d^3k_j}{\omega_j} k_j^\nu k_j^\mu \partial_{\mu} f_{j}\!\left(\BV[k]_j,\BV[x],t\right), \label{st2}
\end{eqnarray}
where $P_{\rm jet}^{{\nu}}$ is the total four-momentum of the jet shower
and $\frac{dP_{\rm jet}^{{\nu}}}{d^3x}$ is its $3$-dimensional space density.
As mentioned above, there are three types of processes, i.e., the collisional energy loss, the transverse momentum broadening, and the medium-induced partonic splitting, contributing to the derivative of the phase-space parton distribution in Eq.~(\ref{st2}).
However, since the total energy and momentum in the jet shower are conserved in the splitting processes, their contribution turns out to vanish by the integration and summation in Eq.~(\ref{st2}).
Therefore, the energy and momentum are exchanged between the jet and the QGP fluid through the processes of first two terms on the right hand side of Eq.~(\ref{eq:jetshower}), i.e., the collisional energy loss and transverse momentum broadening.
We also have to estimate the position of each energetic parton in the full jet. For a parton with energy $\omega_j$ and momentum $\BV[k]_j$ at time $t$, its position is estimated as: $\BV[x]={\BV[x]}_0^{\rm jet}+(\BV[k]_j/\omega_j)t$, where $\BV[x]_0^{\rm jet}$ is the production point of the jet.
Then the source term is obtained as:
\begin{eqnarray}
  \!\!\!\!\!\!\!\!J^{\nu}\!\!\left(x\right)
  &=&
  -\!\sum_j\!\!
  \int \!\!d^3\!k_j
       {k^\nu_j}\!\!
       \left.\frac{d\!f_j \!\left(\BV[k]_j,\!t\right)}{d t}\!\right|_{\rm col.}\!\!
       \delta^{(3)}\!\!\left(\!\BV[x]\!-\!{\BV[x]}_0^{\rm jet}\!\!-\!\frac{\BV[k]_j}{\omega_j}t\!\right)\!\!.
\end{eqnarray}
Here $\left.df_i \!\left(\BV[k],\!t\right)/d t \right|_{\rm col.}$ is the part of the time derivative of the momentum distribution corresponding to the first and second terms on the right hand side in Eq.~(\ref{eq:jetshower}):
\begin{eqnarray}
\!\!\!\!\!\!\!\!\!\!\!\!\!\!\!\!
  \!\!\left.\frac{d{f}_j \!\!\left(
    \omega,k^2_\perp,
    t\right)}{dt}\!\right|_{\rm col.}\!\!\!
  &=&\! \left(\!\hat{e}_j \frac{\partial}{\partial \omega_j}
  \!+\! \frac{1}{4} \hat{q}_j {\nabla_{k_\perp}^2}\!\!\right){f}_j\!(\omega_j, k_{j\perp}^2, t).\label{eq:eqpart}
\end{eqnarray}
Considering the rotational symmetry for the distributions of the shower partons along the jet axis and using the collimated shower approximation ($r = \theta \approx \sin\theta = \frac{k_\perp}{\omega}$), the source term can be written as:
\begin{eqnarray}
  \!\!\!\!\!\!\!\!J^{\nu}\!\!\left(x\right)\!
  &\approx&-
  \frac{1}{2\pi r t^3}\!\!
  \left({x^\nu \!\!- x^\nu_{\rm jet,0}}\right)\!\!\!
  \left.\frac{d E^{\rm jet}}{d t dr}\!\right|_{\rm col.}\!
  \delta\!\left(|\BV[x]\!-\!\BV[x]_0^{\rm jet}\!|\!-\!t\right)\!,\label{eq:stdmom}
\end{eqnarray}
where
\begin{eqnarray}
  \!\!\!\!\!\!\!\!\!\!\!\!
  \left.\frac{d E^{\rm jet}}{d t dr}\!\right|_{\rm col.}\!
  \!\!&=&\!\!\sum_j\!\!
  \int\! \!\!d\omega dk^2_{j\perp}
         {\omega_j}\!\!
         \left.\frac{d{f}_j\!\!\left(
           \omega_j,k^2_{j\perp},
           t\right)}{dt}\!\right|_{\rm col.}\!\!\!
         \!\delta\!\!\left(\!r -\frac{k_{j\perp}} {\omega_j}\!\!\right)\!\!. \label{eq:dmom}
\end{eqnarray}
Here $r$ is defined as $r=\sqrt{(\eta_{\rm s}-\eta_{\rm jet})^2+(\phi'-\phi_{\rm jet})^2}$,
with $\phi'=\arctan[(y-y_0^{\rm jet})/(x-x_0^{\rm jet})]$ denoting the azimuthal angle with respect to the initial position of the jet center $\BV[x]_0^{\rm jet}$, $\eta_{\rm jet}$ and $\phi_{\rm jet}$ the pseudorapidity and azimuthal angle of the jet, and $\BV[x]^{\rm jet}(t)$ the position of the jet center at the time $t$.
Using Eqs.~(\ref{eq:eqpart}), (\ref{eq:stdmom}) and (\ref{eq:dmom}), we can construct the source terms from the energy and transverse momentum distributions of the partons inside the jet shower.
To obtain the source terms
in the relativistic $\tau$-$\eta_{\rm s}$ coordinates for Eq.~(\ref{eq:covariant}),
we perform Lorentz transformation:
\begin{eqnarray}
J^{\bar{\nu}}(\tau, x, y, \eta_s)
&=&
-\frac{dP_{\rm jet}^{\bar{\nu}}}{\tau d\tau dx dy d\eta_s}\nonumber\\
&=&
\Lambda^{\bar\nu}_\mu
J^{\mu}\!\!\left(x\right)
 =
- \Lambda^{\bar\nu}_\mu
 \frac{dP_{\rm jet}^{{\mu}}}{dt d^3x},
\label{ltst}
\end{eqnarray}
where
\begin{eqnarray}
\Lambda^{\bar\nu}_\mu
&=&
\begin{pmatrix}
\cosh\eta_{\rm s}&0&0&-\sinh\eta_{\rm s}
\\
0&1&0&0\\
0&0&1&0\\
-\frac{1}{\tau}\sinh\eta_{\rm s}&0&0&\frac{1}{\tau}\cosh\eta_{\rm s}
\end{pmatrix}.
\label{ltop}
\end{eqnarray}

%%%%%%%%%%%%%%%%%%%%%%%%%%%%%%%%%%%%%%%%%%%%%%%%%%%%%
\subsection{Initial Condition of the Medium}

We assume that the QGP medium is locally thermalized at the proper time $\tau=\tau_0$ (set as $\tau_0 = 0.6$~fm/$c$ in our calculation), and then apply relativistic hydrodynamics to describe its space-time evolution.
We set the initial entropy density distribution in the transverse plane at midrapidity $\eta_{\rm s}=0$ as:
\begin{eqnarray}
\!\!\!\!s_T\left(\BV[x]_{\perp}\right)
	=
	\frac{C}{\tau_0}\,
	\left[
		\frac{\left(1-\alpha\right)}{2}
n^{\scalebox{0.8}{\BV[b]}}_{\rm part}\left(\BV[x]_\perp\right)
+\alpha n^{\scalebox{0.8}{\BV[b]}}_{\rm coll}\left(\BV[x]_\perp\right)
	\right]\!,\label{eq:sT}
	\end{eqnarray}
where $n_{\rm coll}^{\scalebox{0.8}{\BV[b]}}$ and $n_{\rm part}^{\scalebox{0.8}{\BV[b]}}$ are the number densities of nucleon-nucleon binary collisions and participating nucleons generated from the optical Glauber model with impact parameter $\BV[b]$, respectively.
The parameters $C=41.4$ and $\alpha=0.08$ are chosen for Pb$+$Pb collision at 2.76{\rm A~TeV} by fitting the centrality dependence of the multiplicity at midrapidity from the ALICE Collaboration \cite{Hirano:2012yy, Aamodt:2010cz}.
In $\eta_{\rm s}$ direction, we initialize the medium profile using the following function form:
\begin{eqnarray}
H\!\left(\eta_{\rm s} \right)&=&
\exp\left[-\frac{\left(\left|\eta_{\rm s}\right|-\eta_{\rm flat}/2\right)^2} {2\sigma_{\eta}^2} \theta \left(\left|\eta_{\rm s}\right|-\frac{\eta_{\rm flat}}{2} \right) \right].
\label{eq:modelinit}
\end{eqnarray}
The function $H(\eta_s)$ has a shape consisting of a flat region around the mid-rapidity,
like the Bjorken scaling solution \cite{Bjorken:1982qr},
and two halves of a Gaussian connected smoothly to the vacuum at both ends of the flat region.
The parameters $\eta_{\rm flat}=3.8$ and $\sigma_\eta=3.2$ are fitted to reproduce the pseudorapidity distribution for the multiplicity for central Pb$+$Pb collisions at 2.76{\rm A~TeV}.
Finally, the full $3$-dimensional profile of the initial entropy density is given as:
\begin{eqnarray}
s\!\left(\tau_0,\BV[x]_{\perp},\eta_{\rm s}\right)
	&=&
s_T\!\left(\BV[x]_{\perp}\right)
H\!\left(\eta_{\rm s} \right).
\label{eq:modelinit0}
\end{eqnarray}
We further assume that there is no initial flow at $\tau=\tau_0$ in the transverse direction, which means that the radial expansion of the medium is caused solely by the initial pressure gradient incorporated in the initial profile condition of Eq. (\ref{eq:modelinit0}).
For the flow velocity in the longitudinal direction, the space-time rapidity component is initially set to zero: $u^{\eta_{\rm s}}(\tau=\tau_0)=0$ \cite{Bjorken:1982qr}.

In this work, we employ the smooth averaged initial profile of the medium obtained from the optical Glauber model, which does not include the initial geometrical fluctuation of the nucleons and their internal structures in the colliding heavy ions.
This initial state fluctuations in principle can affect both the jet evolution and the medium response to jet quenching; we would like to leave the event-by-event studies including the initial state fluctuations as a future work.

\subsection{Freeze-out}

The hydrodynamic medium response to the energy and momentum deposited from the jets will induce additional flows in the QGP medium.
These jet-induced flows propagate and enhance hadron emissions in directions around the jet axes.
Some of the hadrons produced from jet-induced flows remain in the jet cone after the background subtraction and are counted as part of the full jets.
To obtain the momentum distributions of particles produced from the medium, we use the Cooper-Frye formula \cite{Cooper:1974mv}:
\begin{eqnarray}
E_i\frac{dN_i}{d^3p_i}
=
\frac{g_i}{\left(2\pi\right)^3}\int_{\Sigma}
\frac
{p_i^\mu d\sigma_{\mu}(x)}
{\exp\!\left[{p_i^{\mu}u_{\mu}(x)}/{T(x)}\right]\!\mp 1}, \label{eqn:C-F}
\end{eqnarray}
where $g_i$ is the degeneracy, $\mp$ corresponds to Bose or Fermi distribution for particle species $i$,
and $\Sigma$ is the freeze-out hypersurface.
The freeze-out is chosen to occur at a fixed temperature $T_{\rm FO}$.
Here a typical value $T_{\rm FO}=140\,{\rm MeV}$ (e.g., Ref.~\cite{Luzum:2010ae}) is used. 
For the sake of simplicity, when calculating the particle spectra via the Cooper-Frye formula, we consider only one species of bosons with the same mass as the charged pions.
The contributions from other hadron specicies are taken into account by replacing $g_i$ by the effective degrees of freedom $d_{\rm eff}$ defined as follows:
\begin{eqnarray}
\!\!\!\!\!e_{\rm lat}(T)&=&\frac{d_{\rm eff}(T)}{\left(2\pi\right)^3}\int d^3p
\frac
{\sqrt{m_{\rm \pi^\pm}^2+\BV[p]^2}}
{\exp\!\left[\sqrt{m_{\rm \pi^\pm}^2+\BV[p]^2}/{T}\right]-1},
\end{eqnarray}
where $e_{\rm lat}$ is the energy density obtained from lattice {QCD} calculations \cite{Borsanyi:2013cga}.

The contribution from the jet-induced medium flow to the full jet is obtained by removing the contribution from the background medium without jet propagation:
\begin{eqnarray}
\Delta \frac{dN}{d^3 p}
&=&
\left.\frac{dN}{d^3 p}\right|_{\mbox{\footnotesize w\!/ \!jet}}
-
\left. \frac{dN}{d^3 p}\right|_{\mbox{\footnotesize w\!/\!o \!jet}}.
\label{eq:subtract}
\end{eqnarray}
The contribution $\Delta dN/d^3 p$ is added to that of the jet shower calculated from the transport equations (\ref{eq:jetshower}) to obtain the final state full jet.
For the particles included in Eq.~(\ref{eq:subtract}), we impose the transverse momenta cut $p^{\rm trk, hyd}_T>1~{\rm GeV}/c$ for inclusive jet analyses, and $p^{\rm trk, hyd}_T>0.5~{\rm GeV}/c$ for dijet analyses following the measurements by CMS Collaboration \cite{Chatrchyan:2013kwa, CMS:2015kca}.

\section{Simulations and Results}\label{sec:res}

In this work, we initialize the jet production points in the transverse plane $\eta_{\rm s}=0$ according to the distributions of the binary nucleon-nucleon collisions which are calculated by using the Glauber model \cite{Miller:2007ri}.
Jet spectra and the momentum distributions of the shower partons inside the full jets are obtained via \pythia\:simulation \cite{Sjostrand:2007gs} with \fastjet\, package \cite{Cacciari:2011ma} used for the full jet reconstruction.
Hard jets are assumed to be created at $\tau = 0$, and travel freely until the thermalization proper time of the QGP, $\tau=0.6~{\rm fm}/c$.
Then the jet shower starts to interact with the QGP and evolves according to the transport equations~(\ref{eq:jetshower}).
The jet-medium interaction is turned off when the local temperature of the medium is below $T_c = 160~{\rm MeV}$.
The interaction in the hadronic matter is usually small compared to the QGP phase, and is neglected in this work.

The medium profile at initial time $\tau=\tau_0$ is calculated by employing Eq.~(\ref{eq:modelinit0}) with the impact parameter $\BV[b]=0$ (central collisions).
The evolution of the medium is governed by the ideal hydrodynamic equations with source terms (\ref{eq:eom}).
As the system expands and cools down, it transits from QGP phase to hadronic phase, and finally the freeze-out occurs.
The momentum distributions of hadrons produced from the medium are calculated via the Cooper-Frye formula (\ref{eqn:C-F}).
After the subtraction of the background (without jet), the remaining part ~(\ref{eq:subtract}) contributes to final state full jets.

Hereafter we call the part of jet described by the transport equations~(\ref{eq:jetshower}) as shower part, and that coming from the jet-induced flow via freeze-out as hydro part.
In this study, we mainly focus on the contribution of the hydro part of the full jet.
Detailed studies of the contribution of each medium modification effect on the jet shower part can be found in Ref. \cite{Chang:2016gjp}.
Also since we perform the simulation independently for each single jet shower, the effect of possible interference between the flows induced by multiple jet showers in one event is not included.

  \begin{figure*}[tbh]
   \includegraphics[width=5.8cm,bb=0 0 252 216]{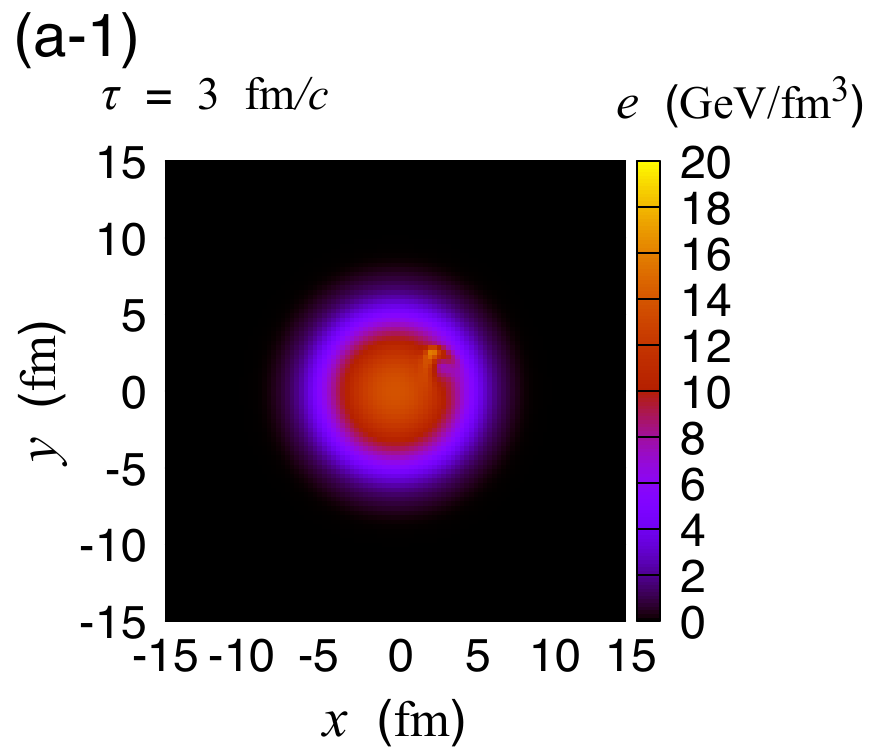}
     \hspace{2pt}
   \includegraphics[width=5.8cm,bb=0 0 252 216]{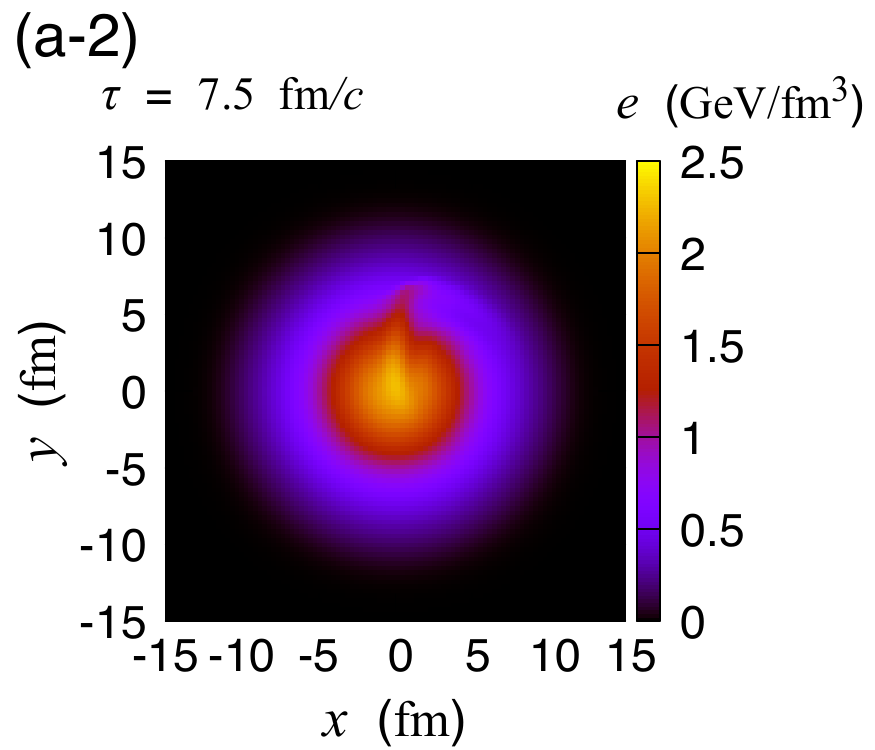}
  \hspace{2pt}
     \includegraphics[width=5.8cm,bb=0 0 252 216]{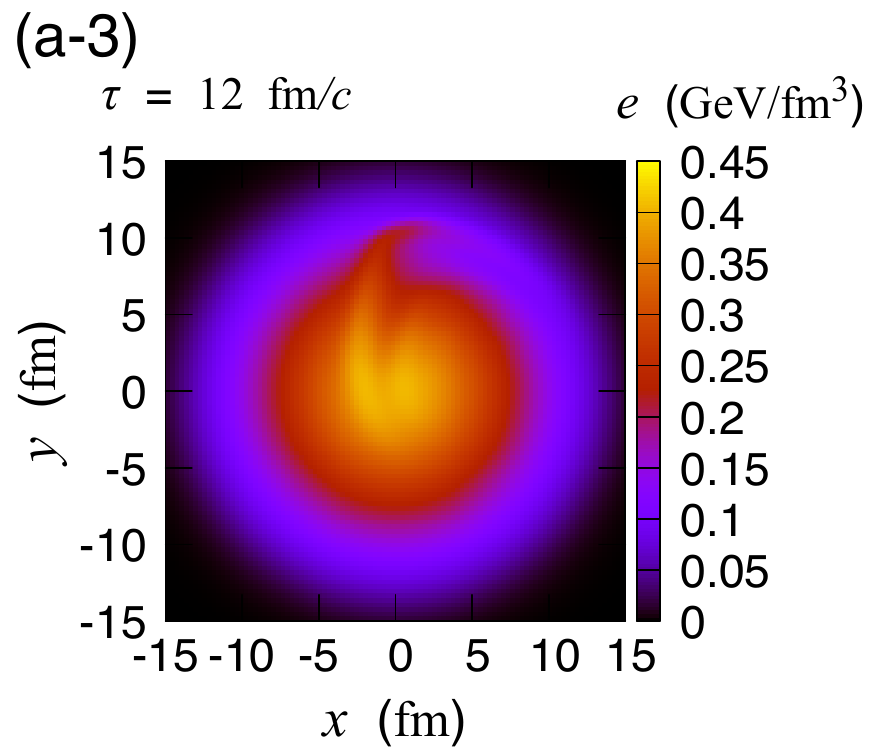}

  \vspace{10pt}

    \includegraphics[width=5.8cm,bb=0 0 252 216]{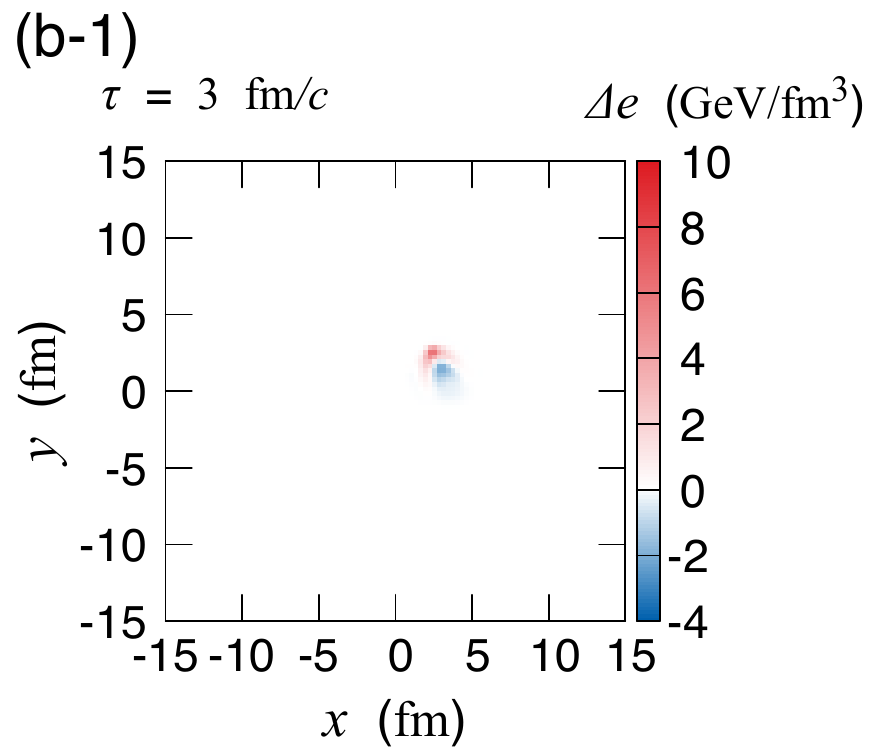}
     \hspace{2pt}
   \includegraphics[width=5.8cm,bb=0 0 252 216]{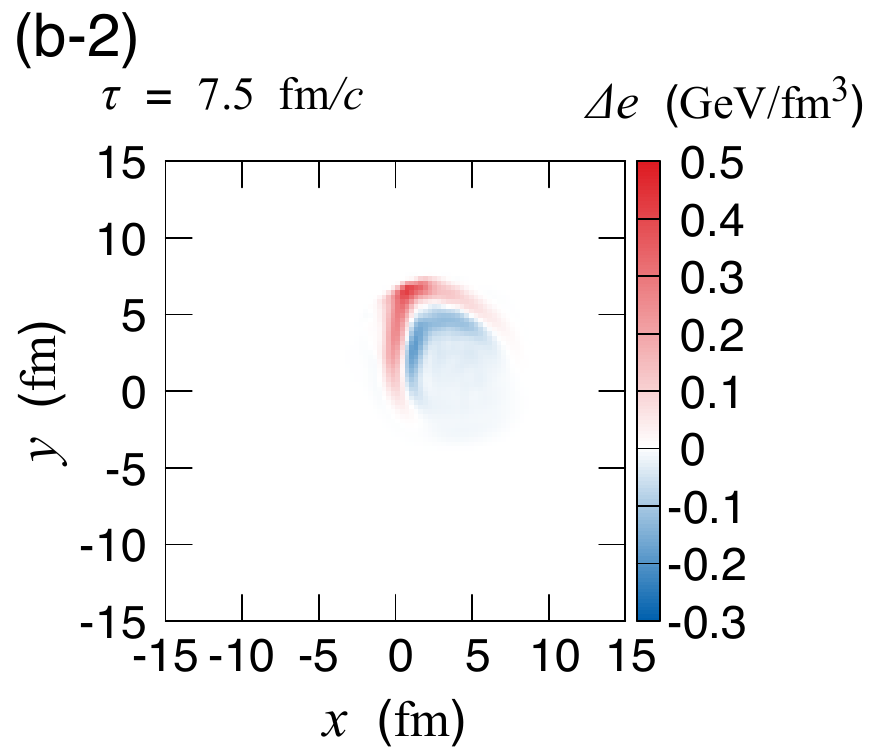}
  \hspace{2pt}
     \includegraphics[width=5.8cm,bb=0 0 252 216]{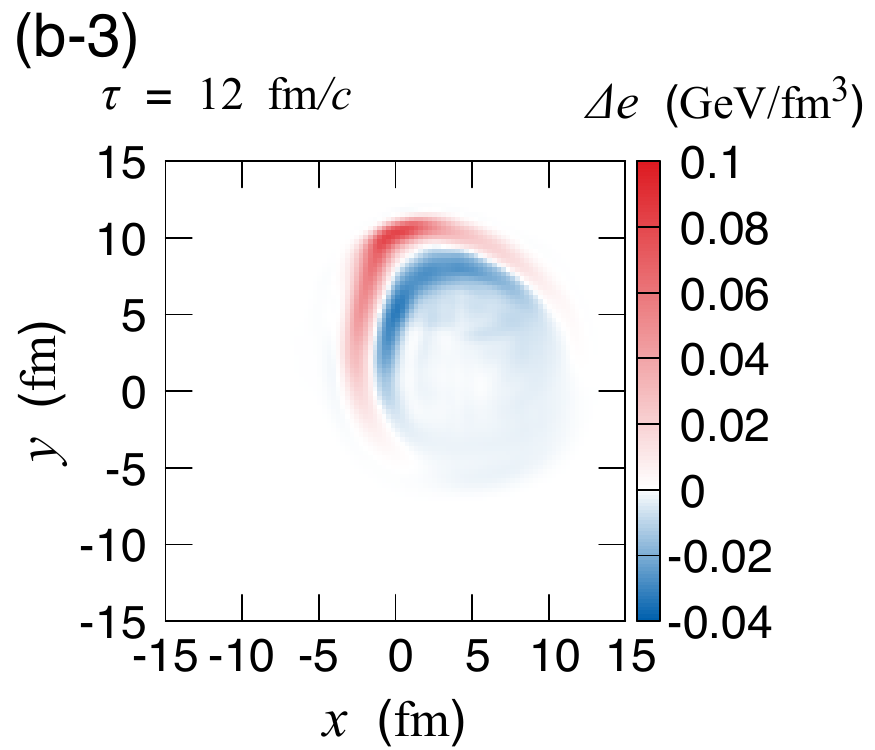}
  \caption{(Color online)
   Energy density distribution of the medium in the transverse plane at midrapidity $\eta_{\rm s}=0$ at different proper times $\tau=3.0$, $7.5$, and $12~{\rm fm}/c$.
   The jet 
   with the initial $p_T^{\rm jet} = 150~{\rm GeV}/c$ 
   is produced at $(x_0^{\rm jet}, y_0^{\rm jet})=(0~{\rm fm}, 6.54~{\rm fm})$ and propagates in the direction of $\phi_p={5\pi}/{8}$.
   The upper panels (a-1), (a-2), and (a-3) show the whole medium energy density, and the lower panels (b-1), (b-2), and (b-3) show the medium energy density subtracted by that in the case without jet propagation.
   }
    \label{fig:evo}
 \end{figure*}

\subsection{Flow Induced by Full Jets}\label{sec:flow}

Figure~\ref{fig:evo} shows the snapshots of the energy density distribution of the medium in the transverse plane at $\eta_{\rm s}=0$ at different proper times for an example event.
For this event, the single jet 
with the initial $p_T^{\rm jet} = 150~{\rm GeV}/c$ 
is produced at $(x_0^{\rm jet}, y_0^{\rm jet})=(0~{\rm fm}, 6.54~{\rm fm})$ and travels in the direction of $\phi_p={5\pi}/{8}$.
The upper panels show the whole energy density of the medium, and the lower panels show the energy density after the subtraction of the energy density in the events without jet propagation.
From these figures, we can see that the V-shaped wave fronts (shown by higher energy density region) are induced by the jet propagation, and develop with time in the medium.
This V-shaped wave front is the Mach cone \cite{Stoecker:2004qu, CasalderreySolana:2004qm, Ruppert:2005uz}, a conical shock wave that appears as an interference of sound waves caused by an object moving faster than the medium sound velocity.
Here the highly collimated jet shower deposits its energy and momentum and induces a Mach cone whose vertex is the center of the jet \cite{Neufeld:2009ep, Qin:2009uh}.
This wave front of the Mach cone carries the energy and momentum, propagates outward and also causes the lower energy density region behind the wave front.
During the propagation, the Mach cone and the radial flow of the medium are pushed and distorted by each other.
One can see that the Mach cone is asymmetrically deformed in this example because the jet travels through the off-central path in the medium.

In this work, we neglect the effect of the finite small shear viscosity of the QGP and model the medium created in relativistic heavy-ion collisions as an ideal (non-viscous) fluid.
The finite viscosities are important for more precise description of the medium evolution and the collective anisotropic flows observed in the final states \cite{Song:2007fn, Schenke:2010rr, Schenke:2011tv,Song:2011qa,Petersen:2011sb, Shen:2011eg}.
It can also affect the shape of the medium response to the jet-deposited energy and momentum, e.g., the Mach cone can be smeared by the finite shear viscosity \cite{Neufeld:2008dx, Neufeld:2010tz, Bouras:2012mh, Bouras:2014rea}.
In our study, we assume the instantaneous thermalization of the energy and momentum deposited by the jet; the finite relaxation time effects may be included in the source terms \cite{Neufeld:2008dx,Iancu:2015uja} (note that the smearing due to the finite grid size in the hydrodynamic simulation mimics some relaxation effect).
Since the relaxation times for the deposited energy and momentum are closely related to the transport coefficients of the QGP, and the inclusion of such effects would provide further information on the QGP's properties, which we would like to leave as a future work.

  \begin{figure*}[tbh]
    \includegraphics[width=8.6cm,bb=0 0 360 360]{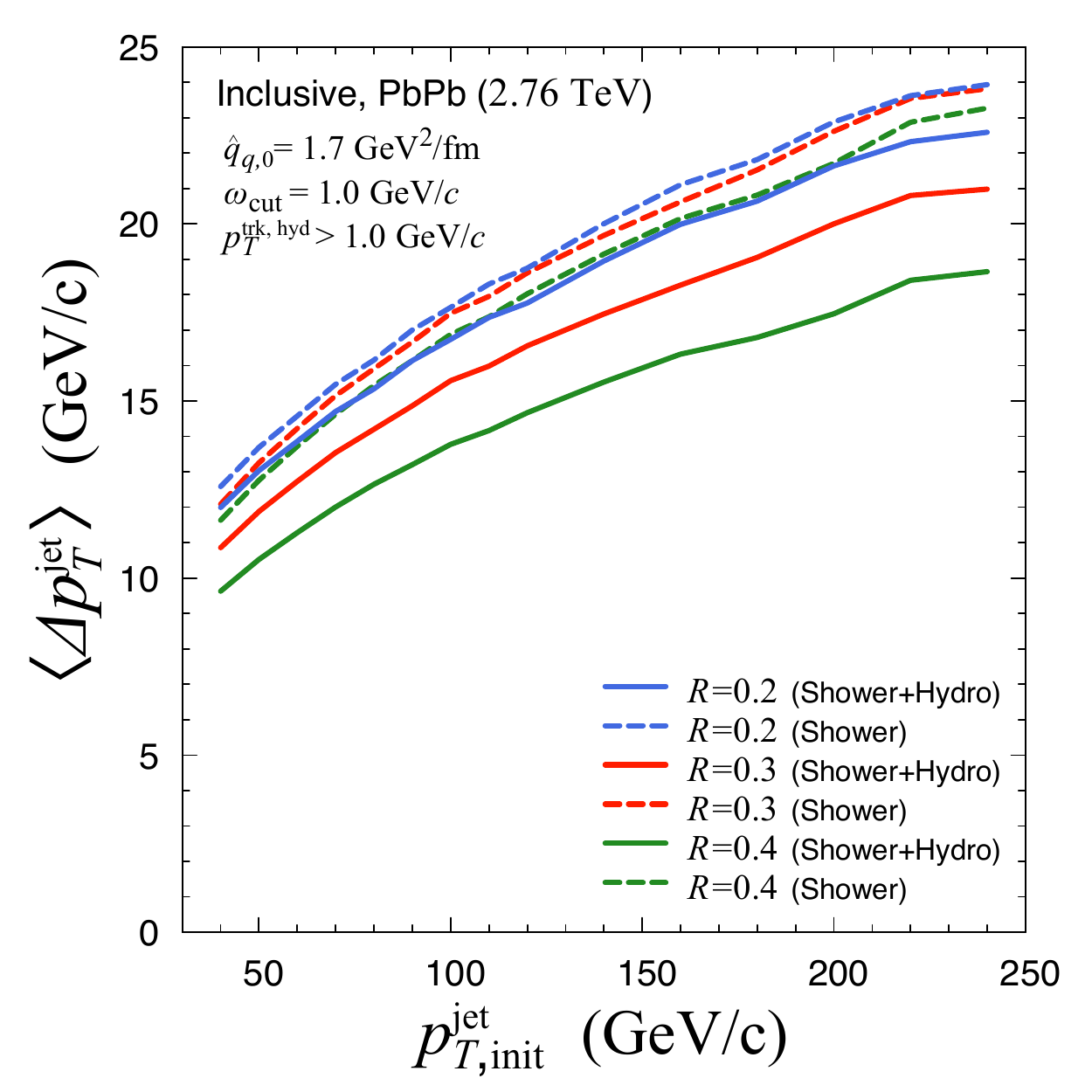}
    \includegraphics[width=8.6cm,bb=0 0 360 360]{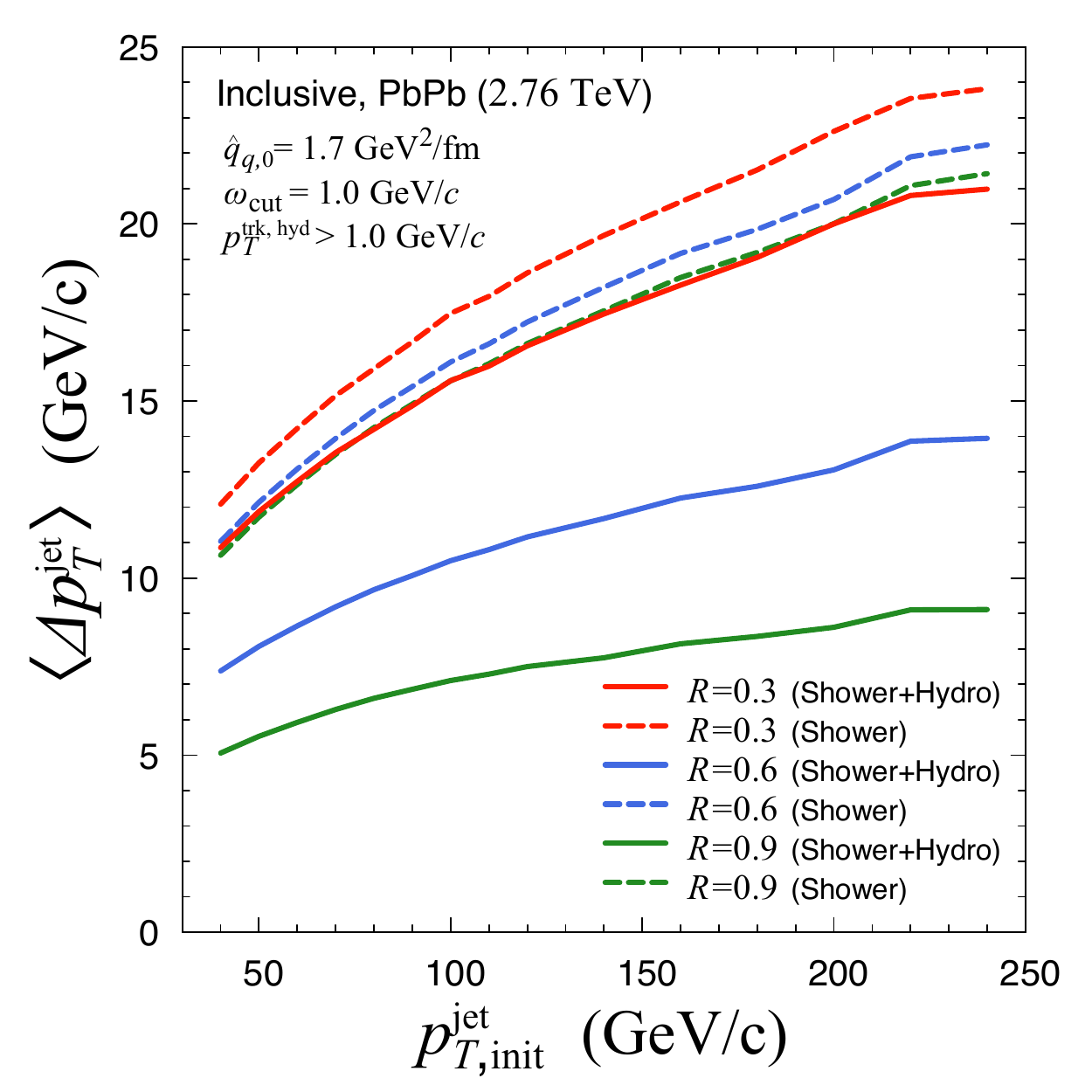}
    \caption{(Color online) %
      Total $p_T$ loss for jets in central Pb+Pb collisions at 2.76{\rm A~TeV} as a function of initial jet $p_T$ for the cone sizes (left panel) $R=0.2$, $0.3$, and $0.4$ and (right panel) $R=0.3$, $0.6$, and $0.9$.
      The solid lines for jets with hydro part, and the dashed lines for jets without hydro part.
    }
    \label{fig:eloss-pt1}
  \end{figure*}

  \begin{figure*}[tbh]	
    \includegraphics[width=8.6cm,bb=0 0 360 360]{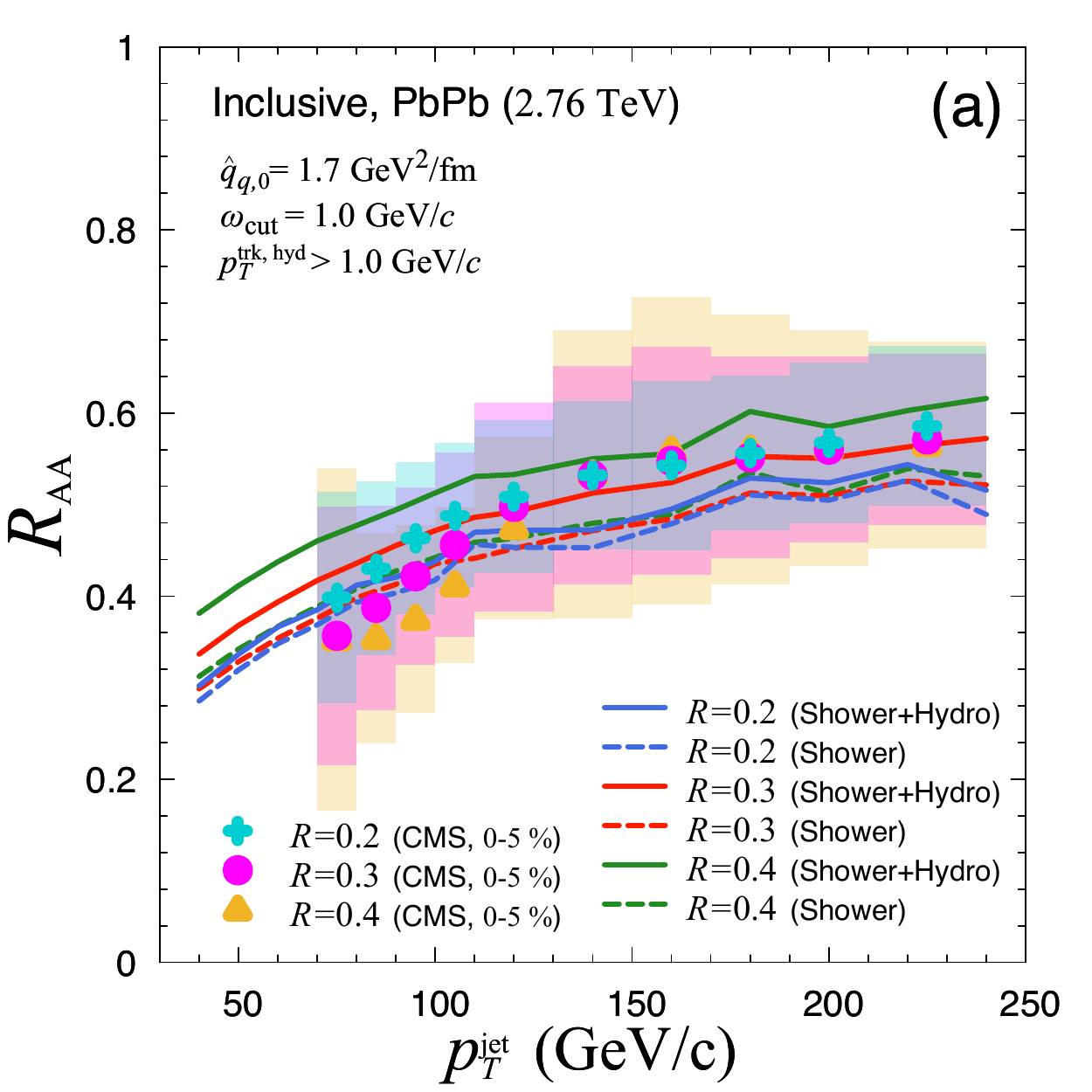}
    \includegraphics[width=8.6cm,bb=0 0 360 360]{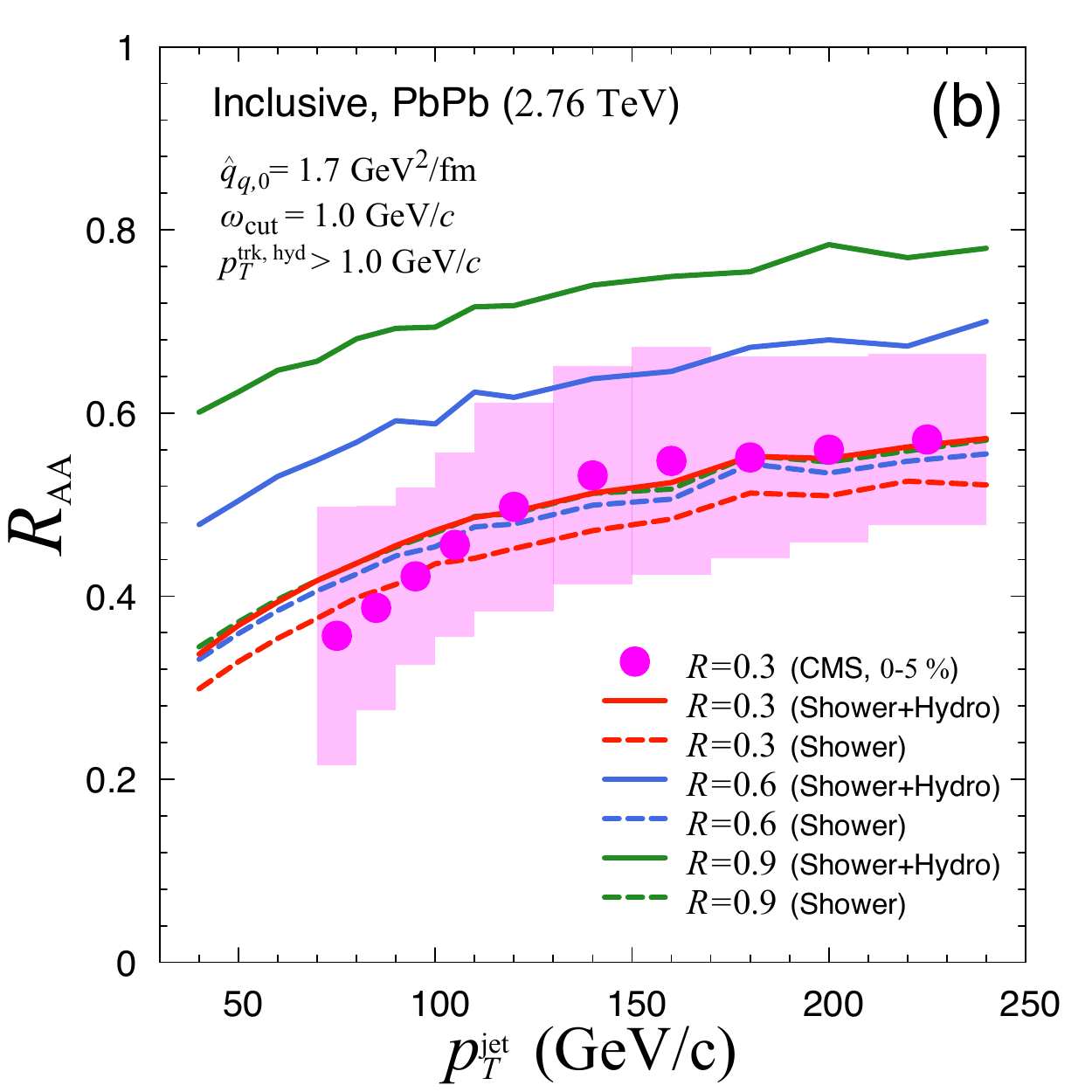}
    \caption{(Color online)
Nuclear modification factor $R_{\rm AA}$ for inclusive jet spectrum
in central Pb+Pb collisions at $2.76{\rm A~TeV}$
with jet cone of the sizes (a) $R=0.2$, $0.3$, and $0.4$, and (b) $R=0.3$, $0.6$, and $0.9$.
      The solid lines are the results for inclusive jets with hydro part, and the dashed lines for without hydro part.
      The turquoise plus markers, the magenta circles, and the orange triangles show
      the experimental data taken from {CMS} Collaboration \cite{Khachatryan:2016jfl}
      for the jet-cone sizes $R=0.2$, $0.3$, and $0.4$, respectively.
      The colored shaded boxes indicate the systematic uncertainties of the same colored data points.}
    \label{fig:raa}
  \end{figure*}

\subsection{Full Jet Energy Loss and Suppression}\label{sec:eloss}

In our framework, the final full jets are contributed from two parts: jet shower part and hydrodynamics response part.
The shower part of the jet loses energy due to three mechanisms: the collisional energy loss and the absorption of the soft partons by the medium, the transverse momentum broadening which kicks the partons out of the jet cone, and the medium-induced radiation outside the jet cone.
The hydro part of the jet comes from the lost energy and momentum from the jet shower which thermalize into the medium and induce connical flow; some of the energy is still inside the jet cone.
Thus the hydro part will partially compensate the energy loss experienced by the jet shower part.
Here we study the effect of jet-induced medium flow on full jet energy loss and full jet suppression.

Figure~\ref{fig:eloss-pt1} shows the mean value of the total energy (transverse momentum) loss for inclusive jets with and without the hydro part contribution as a function of initial full jet transverse momentum.
The left panel shows the results for the jet-cone sizes $R=0.2$, $0.3$, and $0.4$, and the right shows for $R=0.3$, $0.6$, and $0.9$.
One can see the general feature that the amount of the energy loss increases with increasing initial jet transverse momentum while the fractional energy loss decreases.
The total $p^{\rm jet}_T$ loss for the full jets with the inclusion of the hydro part contribution is smaller than that without the contribution from the medium response.
For jets with the cone size $R=0.3$, about $10~\%$ of the lost $p^{\rm jet}_T$ from the jet shower part is recovered by the hydro part.

We can also see the jet cone size dependence of jet energy loss from Figure~\ref{fig:eloss-pt1}.
For the shower part without the hydro part contribution, the jet cone size dependence is rather weak.
This is due to the reason that the shower part of the jet is quite collimated, i.e., most of the energy in the shower part is covered by a narrow jet cone, therefore, jet energy does not change much with increasing jet cone sizes.
On the contrary, jet-induced flow evolves with medium, diffuses, and can spread quite widely around jet axis. As a result, the jet cone size dependence becomes much stronger when adding the hydro part contribution.

The effect of full jet energy loss in the relativistic heavy-ion collisions can be quantified by the measurements of nuclear modification factor $R_{\rm AA}$ for single inclusive jet spectrum, defined as:
\begin{equation}
  \label{eq:raa}
  R_{\rm AA}= \frac{1}{\langle N_{\rm coll}\rangle}\frac{d^2N_{\rm jet}^{\rm AA}/d\eta_{p} dp^{\rm jet}_T}{d^2N_{\rm jet}^{\rm pp}/d\eta_{p} dp^{\rm jet}_T},
\end{equation}
where $\langle N_{\rm coll}\rangle$ is the number of binary nucleon-nucleon collisions averaged over events in a given centrality class, $N_{\rm jet}^{\rm AA}$ is the number of jets in nucleus-nucleus collisions, and $N_{\rm jet}^{\rm pp}$ is that in p+p collisions.
One important result of jet energy loss is that jet $p_T$ spectrum in nucleus-nucleus collisions is shifted to lower $p_T^{\rm jet}$ compared to that in p+p collisions.
Since the jet spectrum is a steeply decreasing function of $p_T^{\rm jet}$, jet $R_{\rm AA}$ will become smaller than unity in high-$p_T^{\rm jet}$ region.

Figure~\ref {fig:raa} shows the nuclear modification factor $R_{\rm AA}$ for single inclusive jets as a function of $p_T^{\rm jet}$ for different jet cone sizes:
the left panel for the jet-cone sizes $R=0.2$, $0.3$, and $0.4$, and the right for $R=0.3$, $0.6$, and $0.9$.
We also compare the results with and without the inclusion of the contribution from the jet-induced flow.
We find that without the hydro part contribution, the jet cone size dependence for jet $R_{\rm AA}$ is very week, which is consistent with the weak dependence for jet energy loss as seen in Figure~\ref{fig:eloss-pt1}.
The inclusion of the contribution from jet-induced flow decreases the total energy loss and thus increase the value of $R_{\rm AA}$; it also increases the jet-cone size dependence of $R_{\rm AA}$.
Our results are comparable with CMS measurements with the jet-cone sizes $R=0.2$, $0.3$, and $0.4$, which show relatively small jet cone size dependence (but with large error bars).

  \begin{figure*}[tbh]
    \includegraphics[width=8.6cm,bb=0 0 360 305]{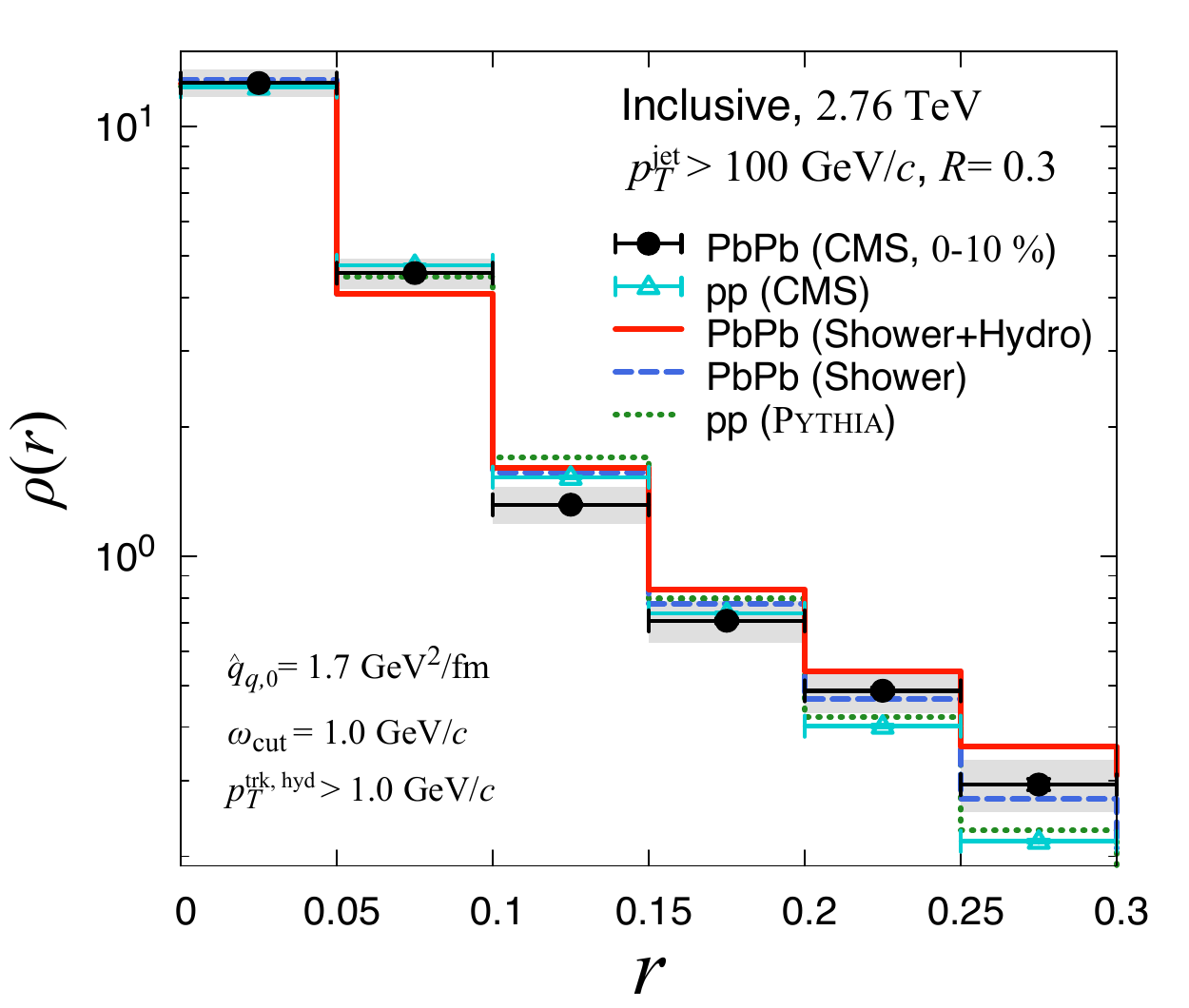}
    \includegraphics[width=8.6cm,bb=0 0 360 305]{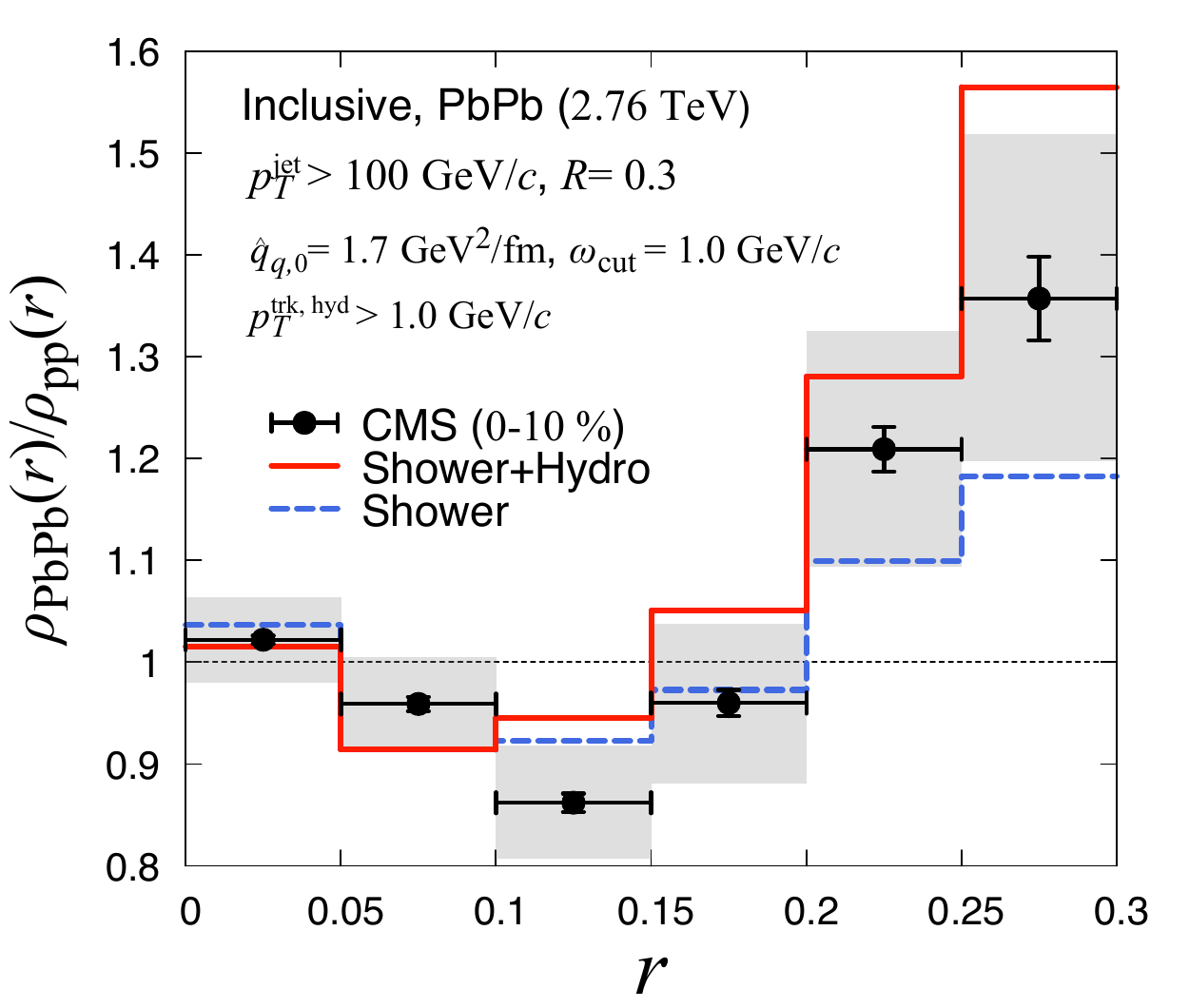}
    \caption{(Color online)
    Left panel: Jet shape function for inclusive jets with $p^{\rm jet}_T>100~{\rm GeV}/c$ in central Pb+Pb and p+p collisions at $2.76{\rm A~TeV}$.
    Right panel: Nuclear modification factor for jet shape function for inclusive jets with $p^{\rm jet}_T>100~{\rm GeV}/c$ in central Pb+Pb collisions.
The solid and dashed lines are results for jets with and without hydro part.
The dotted line is \pythia\:simulation, and the turquoise triangles are data for p+p collisions (left panel).
The black circles are data for Pb+Pb collisions with the shaded boxes for the systematic uncertainties.
Data are taken from {CMS} Collaboration \cite{Chatrchyan:2013kwa}.
}
    \label{fig:shape1}
  \end{figure*}

\begin{figure}[bht]
    \includegraphics[width=8.6cm,bb=0 0 360 305]{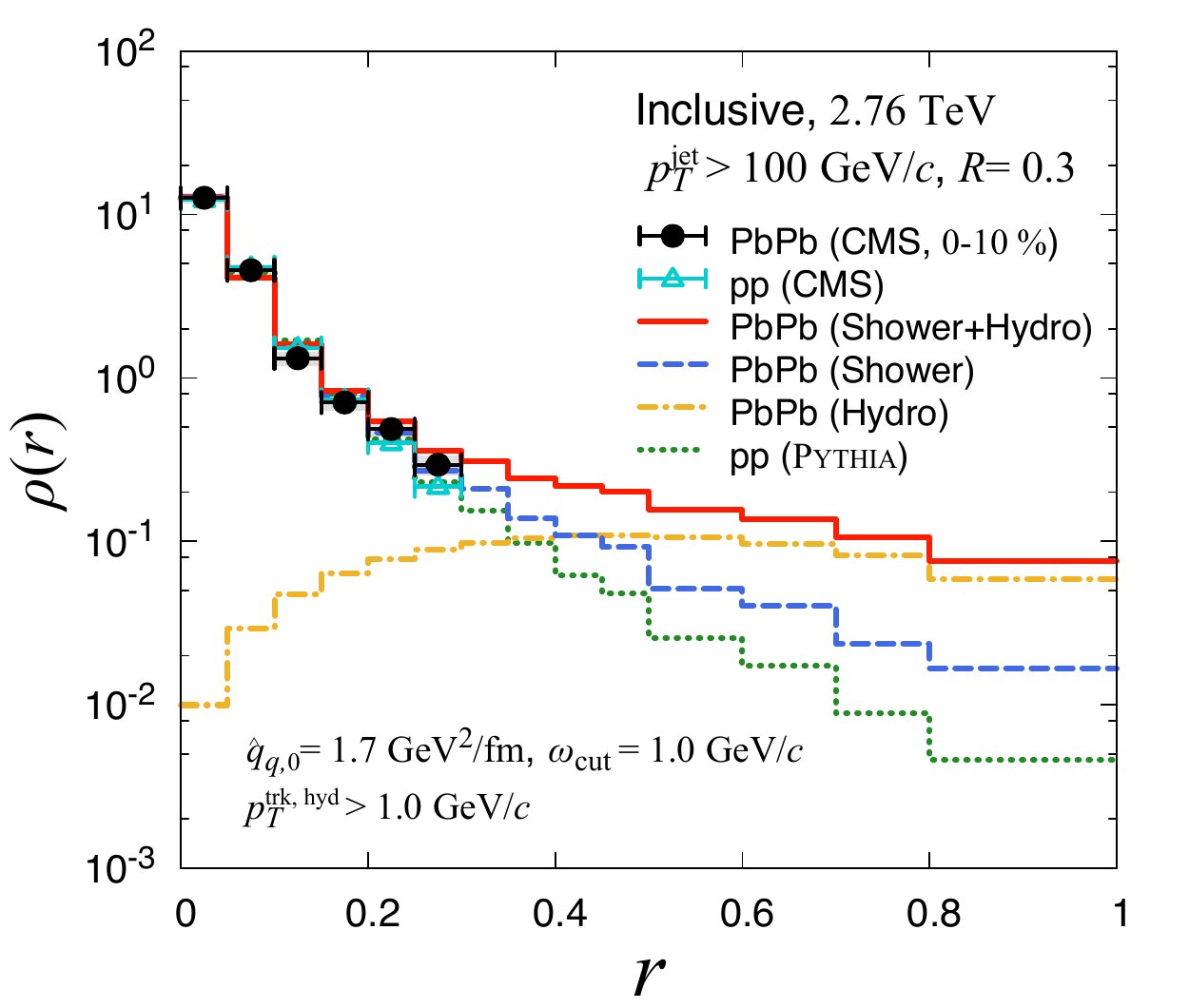}
    \caption{(Color online)
    Jet shape function for single inclusive jets with $p^{\rm jet}_T>100~{\rm GeV}/c$ in central Pb+Pb and p+p collisions at $2.76{\rm A~TeV}$.
    The solid and dashed lines are results for jets with and without hydro part, and the dash-dotted line shows the pure hydro part contribution.
    The dotted line is \pythia\:simulation, and the turquoise triangles are data for p+p collisions.
    The black circles are data for Pb+Pb collisions with the shaded boxes indicating the systematic uncertainties.
    Data are taken from {CMS} Collaboration \cite{Chatrchyan:2013kwa}. }
    \label{fig:shape2}
  \end{figure}

  \begin{figure*}[tbh]
    \includegraphics[width=8.6cm,bb=0 0 360 305]{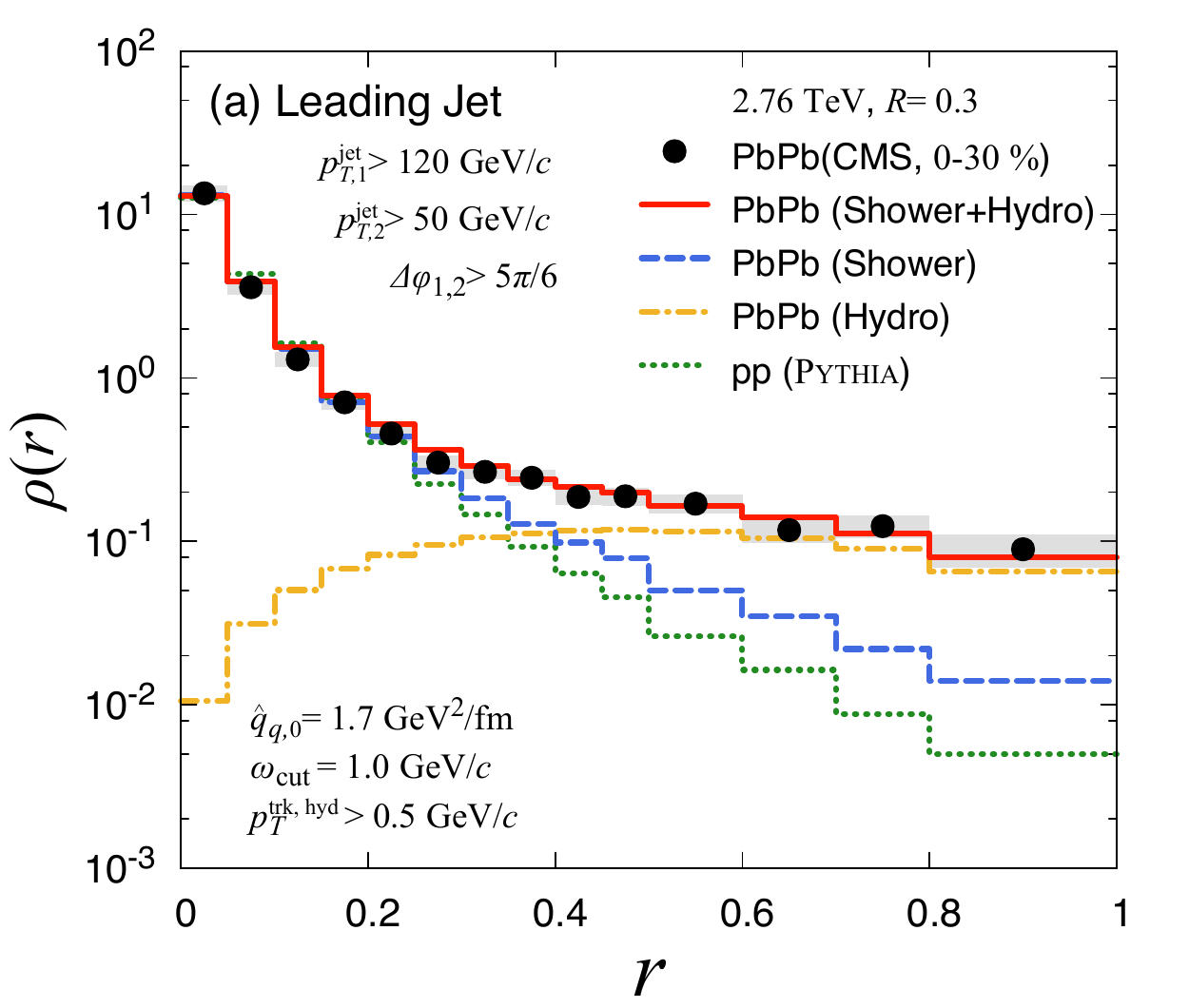}
    \includegraphics[width=8.6cm,bb=0 0 360 305]{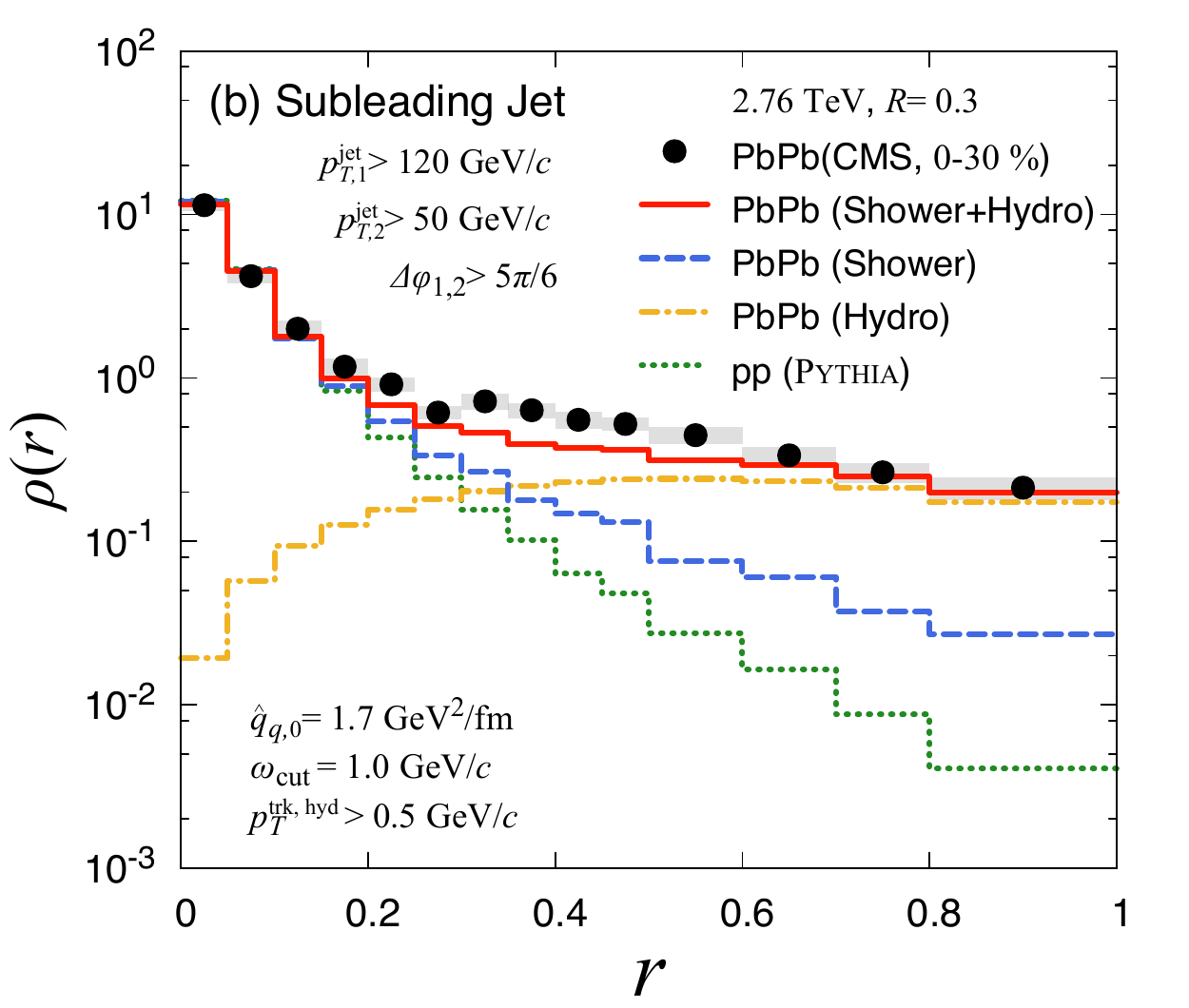}
    \caption{(Color online)
Jet shape function for (a) leading and (b) subleading jets in dijet events in central Pb+Pb collisions and in p+p collisions at $2.76{\rm A~TeV}$.
The leading jet $p^{\rm jet}_{T,1} > 120~{\rm GeV}/c$, the subleading jet with $p^{\rm jet}_{T,2} > 50~{\rm GeV}/c$, and the azimuthal angle between the leading and subleading jets $\Delta\varphi_{1,2}>5\pi/6$.
    The solid and dashed lines are results with and without hydro part contribution in central Pb+Pb collisions, respectively.
    The dashed dotted line shows the pure hydro part contribution.
    The dotted line is result for p+p collisions obtained from \pythia.
    The black circles are data in central ($0$-$30\%$) Pb+Pb collisions from {CMS} Collaboration \cite{CMS:2015kca}, with the shaded boxes indicating the systematic uncertainties.
    }
    \label{fig:dijet}
  \end{figure*}

\subsection{Full Jet Shape Function}\label{sec:shape}

One of the advantages of studying fully reconstructed jets in relativistic heavy-ion collisions is that one may investigate not only the full jet energy loss and suppression, but also their internal structures which provide us the detailed information on how the energy is distributed inside the full jets and how the energy distribution is modified by the interaction with the QCD medium.
Jet shape function describes how the energy inside (and outside) the full jets is distributed in the radial direction (transverse to the jet axis) and is defined as follows:
\begin{eqnarray}
  \label{eq:rho_r}
  \!\!\!\!\!\!\!\!\rho_{\rm jet}(r) &=&
    \frac{1}{N_{\rm jet}}
  \sum_{\rm jet}
  \left[
    \frac {1}{p_T^{\rm jet}}
  \frac{ \sum_{\rm trk\in(r-\delta r/2,r+\delta r/2)} p_T^{\rm trk}}{\delta r}
  \right],
\end{eqnarray}
where $r = \sqrt{(\eta_p - \eta_{\rm jet})^2 + (\phi_p - \phi_{\rm jet})^2}$ is the radial distance of the jet constituents from the jet axis, $\delta r$ is the bin size,
and the sum is taken over all constituents (tracks) of the full jets in the bin at $r$.

The left panel of Figure \ref{fig:shape1} shows our result for the jet shape function inside the jet cone for inclusive jets with $p^{\rm jet}_T>100~{\rm GeV}/c$ and $R=0.3$ in central Pb+Pb collisions and in p+p collisions, compared to the experimental data from CMS Collaboration \cite{Chatrchyan:2013kwa,CMS:2015kca}.
To see the medium effect on the jet shape function more clearly, the nuclear modification factor for the jet shape function $R^\rho_{\rm AA}(r)= \rho_{\rm AA}(r)/\rho_{\rm pp}(r)$ is shown in the right panel of Figure~\ref{fig:shape1}.
We can see that our results (both with and without the contribution from the hydrodynamic response part) show similar nuclear modification pattern for the jet shape function to the experimental data from CMS Collaboration, i.e., little change for small $r$, a dip at $r\sim0.1$ and an enhancement at large $r$.
In other words, the inner hard core of the jet is more collimated while the tail (the outer soft part) of the jet is broadened, in central Pb$+$Pb collisions compared to pp collisions.

The medium modification feature for the shower part of the full jet has been extensively studied in Ref. \cite{Chang:2016gjp} which shows that the collisional energy loss and the thermalization of the soft shower partons (into the medium) make the jet narrower with more collimated hard core, while the transverse momentum broadening and medium-induced radiation transport the energy from the inner to the outer sides of the jet and broaden the tail of the jet shape function.
After the inclusion of the contribution from jet-induced medium flow, the jet shape function at small $r$ is not modified much, but for large $r$ region ($r>0.2$-$0.25$), there is a significant enhancement of the jet broadening effect.
This seems to be quite natural considering the jet cone size dependence of full jet energy loss as seen in Figure~\ref{fig:eloss-pt1}, i.e., the energy loss from the shower part of the jet induces conical flow and medium excitation which evolve with the medium and diffuse to larger angles with respect to the jet axis.

To see more clearly the contribution from the hydrodynamic response part (jet-induced medium flow) to jet broadening effect, we show in Figure~\ref{fig:shape2} the jet shape function $\rho(r)$ for inclusive jets with an extended radial distance $0 < r<1$.
The trigger $p_T$ threshold for the inclusive full jets is set to be $p_T^{\rm jet}> 100~{\rm GeV/}c$.
Here we still use $p_T^{\rm jet}$ defined by the jet-cone size $R=0.3$ as the normalization factor for the jet shape function at $r>R=0.3$, to be consistent with the experimental results from CMS Collaboration \cite{Chatrchyan:2013kwa,CMS:2015kca}. 
The red solid line shows the result for jets with both shower and hydro parts, 
the orange dashed dotted solid line shows the contribution from the hydro part,
and the blue dashed line shows the result for jet without hydro part. 
The green dotted line shows the result from \pythia\  simulation.
As we can see, the shower part of jet shape function is a deep falling function of $r$, while the energy (momentum) from the hydrodynamic response part is a quite flat distribution in a wide range of $r$.
This is because the energy loss from the shower part is carried away by the jet-induced flow which evolves with the medium and diffuses to large distances \cite{Tachibana:2014lja}.
Compared to \pythia\  simulation, the broadening of the shower part of the jet continues to large $r$ by the transverse momentum kicks and medium-induced radiation, but the contribution from the hydro part to the jet shape function is quite flat and finally dominates over the shower part in the region with $r>0.5$.

CMS Collaboration has recently measured the jet shape functions with a wide range of $r$ (up to $r=1$) for both leading and subleading jets in asymmetric dijet events in Pb$+$Pb collisions at 2.76{\rm A~TeV} \cite{Chatrchyan:2013kwa}.
We also perform the calculation for the jet shape functions in dijet events, and the comparison with CMS data is shown in Figure~\ref{fig:dijet}.
In the calculation, we chose dijet events with 
the leading jet with $p^{\rm jet}_{T,1} > 120~{\rm GeV}/c$, 
the subleading jet with $p^{\rm jet}_{T,2} > 50~{\rm GeV}/c$, 
and the azimuthal angle between the leading and subleading jets $\Delta\varphi_{1,2}>5\pi/6$.
In the figure, we do not show the jet shape function for p+p collisions from CMS \cite{Chatrchyan:2013kwa} since the data contain the contamination from the underlying event (and therefore are quite different at large $r$ region as compared to the jet shape function obtained from \pythia\ simulation).
In Pb$+$Pb collisions, such background effect is supposed to be small at large $r$, since the jet shape function is dominated by the shower part at small $r$ and by the jet-induced medium flow at large $r$ (see below).

From Figure~\ref{fig:dijet}(a), we can see that the jet shape function for leading jets in central Pb$+$Pb collisions is quite similar to that for inclusive jets shown in Figure~\ref{fig:shape2}.
The shower part dominates the jet shape function at relatively small $r$ and is broadened by the medium effect starting from $r=0.2-0.3$, while the hydro part starts to dominate the jet shape function at large $r$ region ($r>0.5$).
Our full result on jet shape function with the contributions from both shower and hydro parts reproduces the experimental result quite well throughout the entire $r$ range (up to $r=1$).
For subleading jets as shown in Figure~\ref{fig:dijet}(b), we can see that the jet shape function is much broader than that of leading jets due to larger jet-medium interaction for subleading jets.
As a result, the shower part of the jet shape function is wider and the hydro part contribution is also larger and more widely distributed than that in leading jets.
Our full result also provides a good description of the jet shape function for subleading jets except the middle $r$ region.
The above results clearly show that the hydrodynamic medium response to jet-medium interaction plays an important role in the study of fully reconstructed jets, especially at large $r$ region.
This means that the jet shape function at large angles with respect to the jet direction provides a good opportunity to study the hydrodynamic medium response to jet quenching.

\section{Summary}\label{sec:sum}

In this work, we have studied the nuclear modifications of full jet structures in relativistic heavy-ion collisions with the inclusion of the contribution from the medium exitations induced by the propagating hard jets.
We have formulated a model which consists of a set of transport equations to describe the full jet shower evolution in medium and relativistic ideal hydrodynamic equations with source terms to describe the dynamical evolution of the QGP medium.
The transport equations control the evolutions of energy and transverse momentum distributions of the shower partons within the full jet.
The contribution from the momentum exchange with the medium via
scatterings with medium constituents is taken into account by collisional energy loss and transverse momentum broadening terms. 
The partonic splitting terms account for the contribution of the medium-induced radiations; the rates for the induced splittings were taken from higher-twist jet energy loss formalism.
The local temperature and flow velocity of the medium are embedded in the jet quenching parameter $\hat{q}$ which controls the amplitudes of all the medium modification processes.
The relativistic ideal hydrodynamic equations with source terms determine the space-time evolution of the medium which exchanges the energy and momentum with the propagating jet shower.
The energy and momentum deposited by the jet shower into the medium fluid are included via the source terms, which may be constructed from the solutions of jet shower transport equations based on the energy-momentum conservation for the combined system of the QGP medium and the jet shower.

Based on our coupled jet shower and QGP fluid model, we have performed the simulations of jet events in central Pb+Pb collisions at 2.76{\rm A~TeV}.
The transport equations for jet shower were numerically solved with the initial conditions generated by \pythia.
We kept track of both the energy and transverse momentum distribution of all shower partons in the full jet until the interaction with the QGP ceases.
The relativistic ideal hydrodynamic equations with source terms were numerically solved in the $(3+1)$-dimensional $\tau$-$\eta_{\rm s}$ coordinates, with the initial profile of the medium obtained from the optical Glauber model.
We found that the additional energy density flow can be induced by the jet shower propagation in the medium, and the jet-induced conical flow is pushed and distorted by the medium radial flow (and vice versa) during the propagation.

To study how the hydrodynamic medium response contributes to the full jet observables, we calculated the particles produced from the hydrodynamic medium response by using the Cooper-Frye formula, which are combined with the jet shower part to obtain the final state full jets.
We generated (di)jet events in central Pb+Pb collisions by using \pythia\ for the momentum distribution and the Glauber model for the spatial distribution.
We calculated the total energy ($p_T$) loss of the full jets for different jet-cone sizes and found that the contribution of the hydro part partially compensates the energy loss of the jet shower.
Such compensation effect increases with increasing jet cone size.
As a result, one obtains stronger jet cone-size dependence for the single inclusive jet $R_{\rm AA}$ when taking into account hydrodynamic response contribution.
The effect of jet-induced medium flow on jet shape functions were studied for inclusive jets as well as for dijets in central Pb$+$Pb collisions at 2.76{\rm A~TeV}.
Jet shape functions for leading jets in dijet events and single inclusive jets are quite similar, while the nuclear modification for subleading jets is larger due to more jet-medium interaction.
Our results showed that the particles produced from jet-induced medium flow do not affect very much the jet shape function at small $r$, but significantly enhance the broadening of the jet shape function, and finally dominate at large $r$ region.
Our full results for jet shape functions can reproduce quite well the experimental data from {CMS} Collaboration \cite{CMS:2015kca}, after taking into account both jet shower and hydrodynamic response contributions.
In summary, we have found that jet-induced flow plays significant roles in the study of jet structure in relativistic heavy-ion collisions, especially at large angles with respect to the jet axis; detailed studies of full jet structure at large $r$ should provide us much information about the medium response effect in the process of jet-medium interaction.

\begin{acknowledgements}
The authors would like to thank X.-N. Wang for discussions and  helpful comments.
Also, Y.T. is grateful to Y. Hirono for discussions regarding numerical implementations.
This work is supported in part by the Natural Science Foundation of China (NSFC) under Grant Nos.~11375072 and 11405066, Chinese Ministery of Science and Technology under Grant No. 2014DFG02050, and the Major State Basic Research Development Program in China under Grant No. 2014CB845404.

\end{acknowledgements}

\bibliography{ref.bib}

\end{document}